\documentclass[12pt]{article}
\usepackage[dvips]{graphicx}
\usepackage{latexsym}

\newcommand{\N}		{{\cal N}}
\newcommand{\F}		{{\cal F}}
\newcommand{\turn}	{{\rm turn}}
\newcommand{\dual}	{{\rm Dual}}
\newcommand{\ci}	{{\rm c1}}
\newcommand{\cii}	{{\rm c2}}
\newcommand{\cs}	{{\rm cs}}
\newcommand{\I}		{{\rm I}}

\newcommand{\IIa}	{{\rm IIa}}
\newcommand{\IIb}	{{\rm IIb}}

\newcommand{\eff}	{{\rm eff}}

\newcommand{\crit}	{{\rm c}}

\newcommand{\BH}	{{\rm BH}}
\newcommand{\planck}	{{\rm pl}}
\newcommand{\DW}	{{\rm DW}}
\newcommand{\mini}	{{\rm min}}

\newcommand{\lnear}	{{\begin{array}{c} < \\[-0.9em] \sim \end{array}}}
\newcommand{\gnear}	{{\begin{array}{c} > \\[-0.9em] \sim \end{array}}}

\newcommand{\fig}[1]	{Figure \ref{#1}}

\newcommand{\EQ}[1]	{(\ref{#1})}

\begin{document}

\begin{flushright}
 \begin{minipage}[b]{43mm}
  hep-th/0307295\\
  WIS/19/03-JULY-DPP\\
 \end{minipage}
\end{flushright}

\renewcommand{\thefootnote}{\fnsymbol{footnote}}
\begin{center}
 {\Large\bf
 Spherical Domain Wall formed by Field Dynamics of \\[0.5em]
 Hawking Radiation 
 and Structure Near Horizon
 }\\
 \vspace*{3em}
 {Yukinori Nagatani}\footnote
 {e-mail: yukinori.nagatani@weizmann.ac.il}\\[1.5em]
 {\it Department of Particle Physics,\\
 The Weizmann Institute of Science, Rehovot 76100, Israel}
\end{center}
\vspace*{1em}

\begin{abstract}
 The Hawking radiation in the vacuum of the spontaneous symmetry breaking
 in the gauge-Higgs-Yukawa theory
 is investigated by
 a general relativistic formulation of the the ballistic model.
 The restoration of the symmetry on the horizon
 and the formation of the spherical domain wall
 around the black hole are shown
 even if the Hawking temperature
 is lower than the critical temperature
 of the phase transition in the gauge-Higgs-Yukawa theory.
 When the Hawking temperature is much lower than the critical temperature,
 the domain wall closely near the horizon is formed.
 The wall is formed by the field dynamics rather
 than the thermal phase transition.
\end{abstract}

%

\newpage
\section{INTRODUCTION}\label{intro.sec}

One of the most interestings
in the quantum field theory with the general relativity
is the thermal radiation from black holes,
which is known as the Hawking radiation \cite{Hawking:1975sw,Hawking:1974rv}.
The temperature and the intensity of the radiation
from the Schwarzschild black hole
increase explosively at the final stage of the evaporation
because 
the black hole loses its mass by the radiation
and its temperature is inversely proportional to the mass
\cite{Hawking:1974rv}.
Therefore the particle physics and the field dynamics
including phase transitions around the black hole
are especially interesting.
Several authors discussed that
the heating-up by the Hawking radiation can 
thermalize the neighborhood of the black hole
and the {\it thermal} phase transition around the black hole can arise.
Cline considered the QCD phase transition
around the black hole whose Hawking temperature is greater than
the critical temperature of the quark-gluon-plasma (QGP)
transition \cite{Cline:1996mk}.
He pointed out that some of the gamma ray bursts can be explained
by the radiation from the QGP fireballs
produced by the Hawking radiation from the primordial black holes.
We showed the thermal electroweak (EW) phase transition around the black hole
and the formation of the spherical domain wall
which separates the symmetric-phase-region
from the broken-phase background \cite{Nagatani:1998gv,Nagatani:2001nz}.
We discussed 
several mechanisms of the baryon-number-production
by the spherical domain wall around a black hole
and proposed the cosmological baryogenesis scenario
by the primordial black holes
\cite{Nagatani:1998gv,Nagatani:2001nz,Nagatani:1998rt}.

\begin{figure}
 \begin{center}
  \includegraphics[scale=1.2]{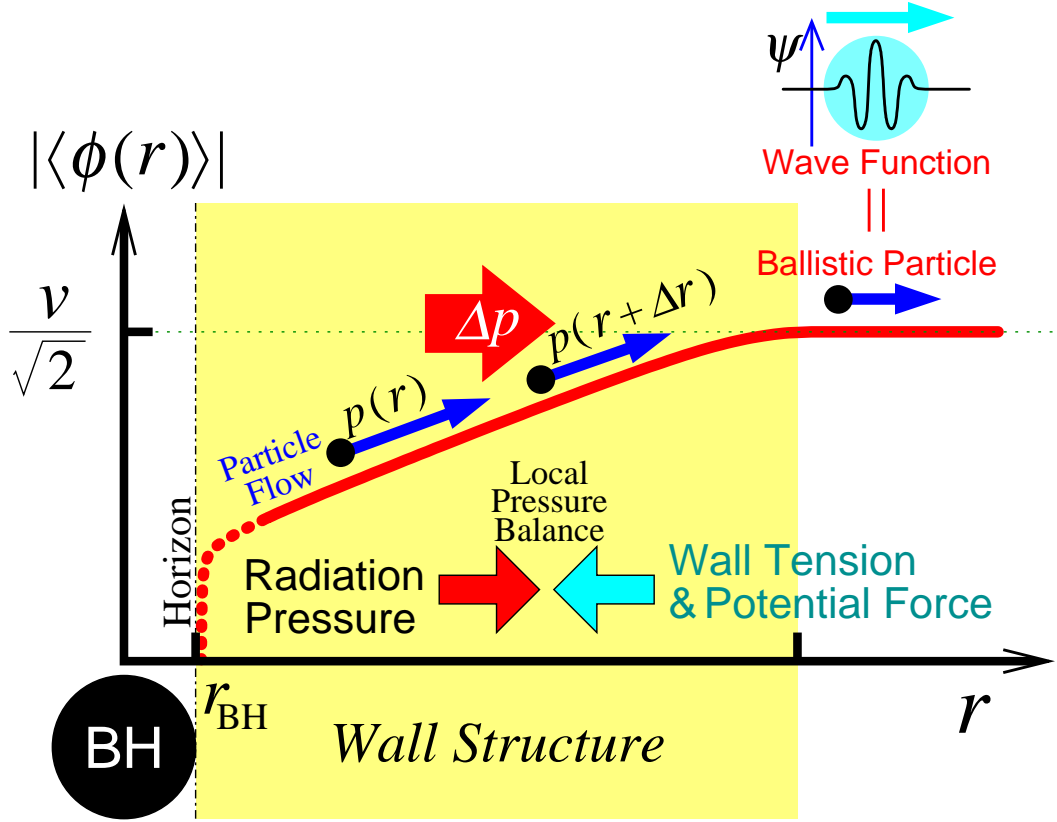}%
 \end{center}
 \caption{%
 The balance of pressures ---
 our basic idea for the wall-formation in the ballistic model.
 The parameter $r$ is the distance from the center of the black hole
 and $|\left<\phi(r)\right>|$ means the absolute value of the Higgs vev.
 Once a wall-structure of Higgs vev (thick curve) is formed,
 mass of the particle is depending on the distance $r$.
 A part of the momentum
 of the Hawking-radiated particle
 $\Delta p = p(r) - p(r+\Delta r)$
 is transfered to the wall-structure
 because the momentum is depending on the distance $r$ in the
 wall-structure: $p(r)=\sqrt{E^2 - m(r)^2}$.
 The transfered momentum causes
 a radiation-pressure acting on the wall.
 A local balance among the radiation-pressure, the wall-tension
 and the potential-force
 can stabilize the wall-structure.
 The dotted thick curve means the wall-structure near the horizon.
 The radiation pressure near the horizon
 is enhanced due to a general relativistic effect
 and forms the wall-structure near the horizon.
 \label{Idea.eps}%
 }%
\end{figure}

The thermal phase transition on some local volume
requires the local thermal equilibrium
which is confirmed by the enough interaction rate
in the considered volume.
The mean free paths of the radiated particles are much longer than
the Schwarzschild radius.
If the thermal phase transition is caused by
heating-up by the Hawking radiation,
the radius of the wall should be much larger than the Schwarzschild radius.
Hence the formation of the wall by the thermal phase transition
needs much higher Hawking-temperature than
the critical temperature of the transition
\cite{Nagatani:1998gv,Nagatani:2001nz}.
Here, it is natural to ask if
the Hawking radiation whose temperature is
similar to the critical temperature of the transition
influences the Higgs vacuum-expectation-value (vev) around the black hole.
In our previous work
\cite{BHBG3,Nagatani:2002cj},
we considered the Hawking radiation with such a temperature
in the gauge-Higgs-Yukawa theory,
e.g., the EW theory or the Grand Unified Theory (GUT).
We proposed {\it the ballistic model}
as an effective description of the system.
By using the ballistic model,
the formation of the wall-structure of the Higgs vev was shown
when the Hawking temperature is equal to or greater than
the energy scale of the gauge-Higgs-Yukawa theory.
The wall-structure is formed by the field dynamics
rather than the thermal phase transition.

The ballistic model is described by both
the action for the Higgs scalar field and
that for the relativistic point particles
as the Hawking-radiated particles from the black hole.
The ballistic description of the radiated particle is valid
because the mean free paths determined by gauge interactions
are much longer than the Schwarzschild radius
\cite{Nagatani:1998gv,Nagatani:2001nz}
and 
the mean wavelength of the radiated particles approximately given by
the Schwarzschild radius is shorter than the length-scale of the
wall-structure
\cite{BHBG3,Parikh:1999mf}.
The basic idea of the wall formation in the ballistic model
is schematically shown in \fig{Idea.eps}.
When once a radius-depending Higgs-vev-structure $|\left<\phi(r)\right>|$,
namely the wall-structure, has been formed,
the Hawking-radiated particles run rush up
the slope on the wall-structure
and they push the wall-structure outside.
Therefore there arises a pressure, named the Hawking-radiation-pressure,
which is acting on the spherical wall-structure to expand its radius.
On the other hand the wall-structure also feels its wall-tension and
the Higgs-potential-force,
which are acting on the wall-structure to shrink its radius.
The balance among these pressures keeps the wall-structure stationary.

In the previous work \cite{BHBG3},
the black hole is assumed as a simple
source of the particles with thermal spectrum
and general relativistic (GR) effects are omitted.
In this paper
we fully consider the GR effects in the ballistic model
and show the formation of the wall-structure
if the Hawking temperature is not only greater than 
the energy-scale of the gauge-Higgs-Yukawa theory
but also smaller than the energy-scale.
We assume that (i) the motion of the particles radiated from the horizon
obeys the geodesic on the Schwarzschild space-time
and (ii) the observer at the infinite distance
detects the disk-image with uniform intensity
as an image of the radiated particles.
By constructing the general relativistic formulation of the ballistic model
we derive the effective Higgs potential including full GR effects,
which determines the Higgs vev structure around the black hole.
The effective potential is depending
on the differential particle-density-distribution $dE \times \N(E,r)$ of
both the particle-position $r$ and the particle-energy $E$.
The analysis of the particle-trajectories radiated from the horizon 
gives us the density-distribution.
The density-distribution $dE \times \N(E,r)$ is depending on
the differential particle-flux
$dE \times d\omega\sin\omega \times f(E,\omega)$ on the horizon
because the initial conditions for the particles
are given by the flux $f(E,\omega)$ on the horizon.
The parameter $\omega$ is a zenith-angle of the particle-radiation
on the horizon.
The elevation-angle of the radiation becomes $(\pi/2 - \omega)$.
The particle-flux for high elevation-angles is derived
by the two assumptions above-mentioned
because particles radiated with the high elevation-angle
can be detected by the observer and their trajectories can be traced.
On the other hand,
the particle-flux for low elevation-angles cannot be determined
by the assumptions
because the particles radiated with a low elevation-angle
return into the horizon and are not detected by the observer.
By the analytic continuation of the particle-flux
from the high elevation-angle to the low elevation-angle,
we can obtain the particle flux for all angle.
Then we can derive the effective potential and
the effective differential equation for the Higgs scalar vev
around the black hole.
The suitable boundary conditions for the differential equation
are as follows;
(i) the Higgs vev $\left<\phi(r)\right>$
approaches the ordinary Higgs vev $\left<\phi\right> = v/\sqrt{2}$
which minimizes the bare Higgs potential $V(\phi)$
when the distance from the black hole
approaches infinity $r\rightarrow\infty$ and
(ii) the Higgs vev should be finite on the horizon.
The field equation is numerically solved with the boundary conditions
and the wall-structure of the Higgs vev around the black hole
is obtained.

Our resultant wall-structure has two interesting properties as follows.
The Higgs wall structure is formed
even if the Hawking temperature is much smaller than
the energy scale of the gauge-Higgs-Yukawa theory.
When we consider such a low temperature black hole,
there arises a wall-structure closely near the horizon.
The Higgs vev on the horizon becomes zero, namely,
the symmetry broken-down spontaneously by the Higgs potential
is restored on the horizon.
These properties are not found in the previous work \cite{BHBG3}
and are essentially caused by the GR effects
on the Hawking-radiated particles.


In Section \ref{ballistic.sec}
the general relativistic formulation of the ballistic model is given
and the general form of the effective Higgs potential $V_\eff$ is obtained.
The effective potential $V_\eff$ is depending on 
the differential density-distribution $dE \times \N(E,r)$
of the radiated particle.
In Section \ref{density1.sec}
various patterns of the particle-trajectories are analyzed
and the form of the density-distribution
$dE \times \N(E,r)$ is obtained
as a functional of the differential flux on the horizon
$dE \times d\omega\sin\omega \times f(E,\omega)$.
In Section \ref{angle.sec}
the flux $dE \times d\omega\sin\omega \times f(E,\omega)$ is discussed.
In Section \ref{density2.sec}
the concrete form of the density-distribution $dE \times \N(E,r)$ is obtained
by using the results of Section \ref{density1.sec}
and Section \ref{angle.sec}.
In Section \ref{formation.sec}
the differential equation for the Higgs vev
and the boundary conditions are discussed and
the formation of the wall-structure is shown.
In section \ref{summary.sec}
we provide a conclusion and discussions.
In Appendix A
another argument of the flux on the horizon
$dE \times d\omega\sin\omega \times f(E,\omega)$
is given.
In Appendix B
the energy-density of the Hawking-radiated particles
around the black hole is calculated.
In Appendix C
analytic solutions near the horizon are discussed.

\section{Ballistic Model with General Relativity}\label{ballistic.sec}

The ballistic model is an effective theory for computing
the structure of the Higgs vacuum expectation value (vev)
around a black hole radiating particles
in the vacuum of the gauge-Higgs-Yukawa theory \cite{BHBG3}.
The Hawking-radiated particle can be regard as ballistic
because the radiated particle has
a longer mean free path and has a shorter wavelength
than the radius of the wall-structure
which we will discuss later.
The ballistic particle is identified with
a relativistic point particle with interactions.
In the ballistic model
the Higgs scalar field is separated into the Higgs vev and
the ballistic Higgs particles \cite{BHBG3}.
The Higgs vev is depending on the position
but is not depending on the time.
The ballistic Higgs particle describes
the Higgs propagating mode in the background of the Higgs vev.
The ballistic model consists of
the time-independent Higgs scalar vev $\phi(x)$
and a set of the trajectories of the ballistic particles $\{y_i^\mu(s)\}$,
where $i$ denotes an index of each particle
and $s$ is a parameter of the trajectory.
The set of the ballistic particles $\{y_i^\mu(s)\}$ contains
all radiated particles around the black hole,
which also includes the ballistic Higgs particles.
The fundamental action for the ballistic model is given by
a combination of an action for the Higgs scalar vev $\phi(x)$
with a Higgs potential $V(\phi)$ and
an action for the trajectories $\{y_i^\mu(s)\}$ of
relativistic point particles as the ballistic particles:
\begin{eqnarray}
 S[\phi,y] &=& \int d^4x \sqrt{g} \,
  \left[\,\rule{0mm}{4mm}
   g^{\mu\nu} \partial_\mu\phi \partial_\nu\phi - V(\phi)
  \,\right]
   \;-\;
   \sum_i \int ds \:
   Y_i |\phi(y_i)| \sqrt{g_{\mu\nu}\dot{y}_i^\mu\dot{y}_i^\nu},
   \label{action1}
\end{eqnarray}
where we have defined $\dot{y}_i^\mu := dy_i^\mu / ds$
and the summation is performed over all ballistic particles.
The background space-time is given by the Schwarzschild metric:
\begin{eqnarray}
 ds^2 &=&
  \:+\: F(r) dt^2 \:-\: F^{-1}(r) dr^2 \:-\:
  r^2 d\theta^2 \:-\: r^2 \sin^2\theta d\varphi^2,
  \label{metric}
\end{eqnarray}
with the Schwarzschild factor:
\begin{eqnarray}
 F(r) &:=& 1 - \frac{r_\BH}{r}.
\end{eqnarray}
The bare Higgs potential in the action is given by the double-well form:
\begin{eqnarray}
 V(\phi)
  &=&  -\frac{1}{2}\mu^2 \phi^2
  \;+\; \frac{1}{2}\frac{\mu^2}{v^2} \, \phi^4,
  \label{bare-pot}
\end{eqnarray}
which has a minimum at $|\phi| = v/\sqrt{2}$
and the constant $\mu^2 > 0$ is the Higgs mass,
therefore, the Higgs vev without the Hawking radiation
is given by $|\left<\phi\right>| = v/\sqrt{2}$.
We have defined an extended-Yukawa-coupling-constant
$Y_i$ for the point particle $i$.
The mass of the particle $i$ is proportional to the Higgs vev
and is given by
\begin{eqnarray}
 m_i &=& Y_i |\left<\phi\right>|,
  \label{Mass-Relation.eq}
\end{eqnarray}
where the mass relation in the gauge-Higgs-Yukawa theory is reproduced.
Especially mass of the particle $i$ in the vacuum is given by
$m_i = Y_i |\left<\phi\right>| = Y_i v / \sqrt{2}$.
The ballistic Higgs particles does not obey the relation \EQ{Mass-Relation.eq},
however, by choosing 
\begin{eqnarray}
 Y_{\rm Higgs} &=& \sqrt{3} \frac{\mu}{v}
\end{eqnarray}
as the extended-Yukawa-coupling-constant for the ballistic Higgs particles,
the action \EQ{action1} results in the correct effective potential
(see Appendix in \cite{BHBG3}).
The formulation presented above
is a simple general relativistic extension
of the original ballistic model in the previous work \cite{BHBG3}.

By a gauge fixing $ds = dt = y_i^t$
and employing the Schwarzschild coordinate
$(t,r,\theta,\varphi)$, the action becomes
\begin{eqnarray}
 S[\phi,y]
  &=& \int dt dr d\theta d\varphi \; r^2 \sin\theta \nonumber\\
  && \times
   \Biggl[
   F^{-1} \dot{\phi}^2
   - F (\partial_r \phi)^2
   - r^{-2} (\partial_\theta \phi)^2
   - r^{-2} \sin^{-2}\theta (\partial_\varphi \phi)^2
   - V(\phi)
   \nonumber\\
  && \;-\;
  \sum_i \; Y_i \: |\phi(t,r,\theta,\varphi)| \; \frac{1}{\gamma_i} \;
  \frac{ \delta(r-y^r) \:
         \delta(\theta-y^\theta) \:
	 \delta(\varphi-y^\varphi) }
       {r^2 \sin\theta}\Biggr],
\end{eqnarray}
where we have defined a general relativistic gamma factor for particle
$y_i$ as
\begin{eqnarray}
 \gamma_i =
  \left[  F
        - F^{-1} (\dot{y}_i^r)^2
	- r^2 (\dot{y}_i^\theta)
	- r^2 \sin \theta (\dot{y}_i^\varphi)^2 
  \right]^{-1/2}.
\end{eqnarray}
We assume spherical symmetry of the Higgs vev $\phi(r)$
due to the spherical symmetry of the metric and the Hawking radiation.
The equation of motion for the Higgs vev
with the condition $\partial_t\phi=0$ becomes
\begin{eqnarray}
  \triangle \phi
   &=&
  \frac{1}{2} \frac{\partial V}{\partial\phi}
  \;+\;
  \frac{1}{2}
  \frac{\partial |\phi|}{\partial\phi} \sum_i Y_i \;
  \frac{1}{\gamma_i} \;
  \frac{ \delta(r-y_i^r) \:
         \delta(\theta-y_i^\theta) \:
	 \delta(\varphi-y_i^\varphi) }
       {r^2 \sin\theta},
  \label{EOM-phi}
\end{eqnarray}
where we have defined a Laplacian with spherical symmetry
\begin{eqnarray}
 \triangle\phi &:=&
  \;+\; \frac{1}{r^2} \partial_r \left( r^2 F \partial_r \phi \right).
\end{eqnarray}
Because 
the metric and the Higgs vev as a background of the ballistic particles
have an invariance of the time-evolution and have spherical symmetry,
there are two constants of motion for each ballistic particle,
namely, the energy $E_i$ and the angular momentum $L_i$.
By using spherical symmetry
we select the coordinate system individually
for each particle trajectory to keep $y^\theta_i=0$.
The motion of each particle
is described on each plane $(y_i^r,y_i^\varphi)$.
By using constants of the motion,
the equation of motion for the ballistic particles becomes
\begin{eqnarray}
 E_i &=& Y_i |\phi(y^r_i)| \, \gamma_i(t) \; F(y^r_i),
  \label{energy-def}\\
 L_i &=& Y_i |\phi(y^r_i)| \, \gamma_i(t) \; r^2 \, \dot{y}^\varphi
  \label{angular-momentum-def}.
\end{eqnarray}
By substituting \EQ{energy-def} into \EQ{EOM-phi} and
using property of the Dirac delta function,
we obtain
\begin{eqnarray}
  \triangle \phi &=&
  \frac{1}{2}
  \frac{\partial V}{\partial\phi}
  \;+\;
  \frac{1}{2}
  \frac{\partial |\phi|}{\partial\phi} \sum_i Y_i^2 \;
  F(r) \frac{|\phi(r)|}{E_i} \;
  \frac{ \delta(r-y_i^r) \:
         \delta(\theta-y_i^\theta) \:
	 \delta(\varphi-y_i^\varphi) }
       {r^2 \sin\theta}.
  \label{EOM-phi-2}
\end{eqnarray}
We define the effective potential as
\begin{eqnarray}
 V_\eff(\phi,y) &:=&
  V(\phi)
  \;+\;
  \frac{1}{2}
  \phi^2 \, F(r) \sum_i Y_i^2 \frac{1}{E_i}
  \frac{ \delta(r-y_i^r) \:
	 \delta(\theta-y_i^\theta) \:
	 \delta(\varphi-y_i^\varphi) }
  {r^2 \sin\theta},
  \label{Effective-Potential-1}
\end{eqnarray}
then the equation of motion for the Higgs vev becomes
\begin{eqnarray}
  \triangle \phi &=&
  \frac{1}{2} \frac{\partial V_\eff}{\partial\phi}.
  \label{EOM-phi-3}
\end{eqnarray}

Instead of considering the effect of each ballistic particle,
we will statistically treat the group of the particle
of the same kind.
We adopt a differential distribution of particle-number-density
$dE \times \N_{f}(E,r)$ for particle species $f$ with energy $E$
at the position $r$.
The density-distribution should has the spherical symmetry
because of the absence of a special direction of the Hawking radiation.
The effective potential \EQ{Effective-Potential-1} becomes
\begin{eqnarray}
 V_\eff(\phi,r)
  &=& V(\phi)
  \;+\; \frac{1}{2}
  \phi^2 \, F(r) \sum_f Y_f^2 \int \frac{dE}{E} \N_f(E,r),
  \label{Effective-Potential-2}
\end{eqnarray}
where the summation in the effective potential \EQ{Effective-Potential-2}
is performed over
all particle-species $\{f\}$ in the gauge-Higgs-Yukawa theory
and $Y_f$ is the extended-Yukawa-coupling-constant
for a particle-species $f$.
We rewrite the effective potential \EQ{Effective-Potential-2} as
\begin{eqnarray}
 V_\eff(\phi, r; \N) &=&
    +   \frac{1}{2} \, \mu_\eff^2(r; \N)  \, \phi^2
  \;+\; \frac{1}{2} \frac{\mu^2}{v^2} \, \phi^4,
 \label{Effective-Potential-3}
\end{eqnarray}
where we have defined the effective Higgs mass as
\begin{eqnarray}
 \mu_\eff^2(r; \N)
  &=& -\mu^2
  \;+\; F(r) \sum_f Y_f^2 \int \frac{dE}{E} \N_f(E,r).
  \label{mu_eff}
\end{eqnarray}
The effective Higgs mass \EQ{mu_eff}
which governs the effective potential \EQ{Effective-Potential-3}
is a functional of the density-distribution $\N_f(E,r)$.
Therefore we have found the Higgs vev structure around the black hole
is determined by the density-distribution
of the Hawking-radiated particles.

\section{Density of the Ballistic Particles I}\label{density1.sec}

\begin{figure}
 \begin{center}
  \includegraphics[scale=1.2]{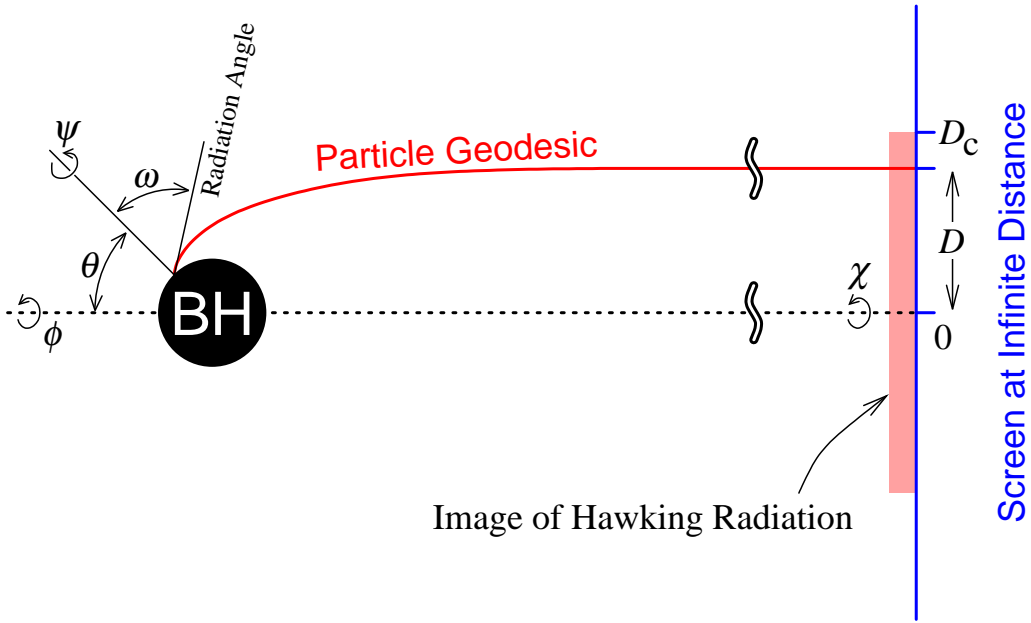}%
 \end{center}
 \caption{%
 The configuration of our setup ---
 the coordinate $(\theta,\phi)$ of the horizon,
 the radiation-angle $(\omega,\psi)$ on the horizon
 and the screen at the infinite distance.
 $\omega$ is the zenith-angle and $\psi$ is the azimuthal-angle
 of the radiation on the horizon.
 The elevation-angle of the radiation becomes $(\pi/2 - \omega)$.
 The screen is parameterized by the polar coordinate $(D, \chi)$,
 where $D$ is the distance from the center of the screen
 and $\chi$ is the angle.
 $D$ is identified as the impact-parameter of the particle-geodesic.
 The observer at the infinite distance finds an image of a disk
 as a detection of the Hawking radiation.
 $D_c$ is the radius of the disk-image,
 which is depending on the ratio of
 the particle-energy $E$ to the particle-mass $m$.
 In this figure the $\phi$-axis and $\chi$-axis correspond,
 however, this correspondence is dispensable.
 The strength of the Hawking radiation is independent of $(\theta,\phi)$
 and of $\psi$ due to the spherical symmetry of the black hole
 and of the Hawking radiation.
 The disk-image is independent of $\chi$ because of the spherical symmetry.
 Each value of $D$ $(< D_c)$ is one-to-one corresponding to $\omega$
 by the particle-geodesic when the particle-energy $E$ ($>m$) is given.
 \label{ImpactParam.eps}%
 }%
\end{figure}

In this section we calculate the density-distribution $\N_f(E,r)$ 
of the ballistic particles radiated from the black hole.
To determine the density-distribution
we assume as follows:
\begin{enumerate}
 \item[(i)]  The particles are radiated from the horizon and
	     obey the equations of the motion
	     in \EQ{energy-def} and \EQ{angular-momentum-def}
	     on the Schwarzschild space-time \EQ{metric}.
\end{enumerate}
The configuration of our setup
is schematically shown in \fig{ImpactParam.eps}.
A part of radiated particles reach infinite distance.
The initial condition for the particle is given by
the position $(\theta,\varphi)$ where the particle radiate on the horizon,
the radiation-angle $(\omega,\psi)$ on the position
and the energy $E$ which is defined in \EQ{energy-def}.
$\omega$ is a zenith-angle defined for $0\leq\omega<\pi/2$
and $\psi$ is a azimuthal angle defined for $0\leq\psi<2\pi$.
The zenith-angle $\omega$ is defined by
the angle which the trajectory of the particle
and the perpendicular of the horizon make on the horizon.
The zenith-angle $\omega$ is defined on the Schwarzschild coordinate.
The elevation-angle of the radiation becomes $(\pi/2 - \omega)$.

We put the differential number-flux of the Hawking-radiated particles
per an unit area on the horizon with the Schwarzschild coordinate
\begin{eqnarray}
 d\F_f &=&
 f_f (E,\omega) \times dE \times d\omega \sin\omega \; d\psi
  \label{Diff-Flux}
\end{eqnarray}
which depends on the energy $E$ of the particle and
on the zenith-angle $\omega$ of the radiation.
The flux \EQ{Diff-Flux} is independent of
the position $(\theta,\phi)$ and of the azimuthal angle $\psi$
due to the spherical symmetry.
The form of the particle-trajectory is
determined by the particle-energy $E$
and zenith-angle $\omega$ of the radiation as the initial conditions,
and is also determined by the mass of the particle, i.e.,
the product of the extended-Yukawa-constant $Y_f$
and the Higgs vev $|\phi(r)|$ as the back ground.
Then the density-distribution $\N_f(E,r)$
is obtained by the analysis of the particle-trajectories
and is depending on the flux $f_f (E,\omega)$.

\subsection{Classification of Particle Trajectory}

We rewrite the motion-equations
\EQ{energy-def} and \EQ{angular-momentum-def}
for the ballistic particle
into the differential equations:
\begin{eqnarray}
 (\dot{y}^r)^2
  &=& F^2 \:
  \left\{
   1
   \;-\; \left(\frac{L}{E}\right)^2 \frac{F}{(y^r)^2}
   \;-\;  \left(\frac{Y|\phi(y^r)|}{E}\right)^2 F
  \right\},\label{eom-r-1}\\
 \dot{y}^\varphi &=& \frac{L}{E} \frac{F}{(y^r)^2}.
  \label{eom-varphi-1}
\end{eqnarray}
We will abbreviate the particle-index $i$ or flavor-index $f$
in equations for simplicity.
By combining the equations \EQ{eom-r-1} and \EQ{eom-varphi-1}
a relation for the zenith-angle $\omega$ is obtained as
\begin{eqnarray}
 \tan \omega
  &=&
  r_\BH \left.\frac{\partial y^\varphi}{\partial y^r}\right|_{y^r = r_\BH}
  \;=\;
  \frac{L}{E} \frac{1}{r_\BH}.
  \label{tanOmegaAndLE}
\end{eqnarray}
By using the relation \EQ{tanOmegaAndLE}
we employ the zenith-angle $\omega$ as a constant of
the particle-motion instead of the angular momentum $L$.
We approximate by
\begin{eqnarray}
 Y|\phi(r)| &\simeq& m
\end{eqnarray}
in the motion-equation \EQ{eom-r-1} for simplifying the analysis,
where $m$ is the mass of the particle
in the ordinary vacuum $\phi=v/\sqrt{2}$\footnote{
This approximation is available
because the form of the final results $\phi(r)$ in \fig{WallForm.eps}
are approximately flat.}.
By this approximation 
the backreaction from the Higgs-vev-structure
into the particle-trajectories is ignored and
the only influence of the radiated particles over the Higgs vev
is considered.
The equations of motion for the ballistic particle
in \EQ{eom-r-1} and \EQ{eom-varphi-1} becomes
\begin{eqnarray}
 (\dot{y}^r)^2
  &=& -W_\eff(y),\label{eom-r-2}\\
 \dot{y}^\phi &=& \tan\omega \: \frac{r_\BH}{(y^r)^2} \:F,
  \label{eom-varphi-2}
\end{eqnarray}
where we have defined an effective potential for the particle-motion
\begin{eqnarray}
 W_\eff(y) &:=&
  - F^2 \:
  \left\{
    1 \;-\; \tan^2 \omega \, \left(\frac{r_\BH}{y^r}\right)^2 \, F
      \;-\; \left(\frac{m}{E}\right)^2 F
  \right\}.
  \label{Particle-Effective-Potential-1}
\end{eqnarray}
The effective potential \EQ{Particle-Effective-Potential-1}
is parameterized by two constants $(m/E, \omega)$.

\begin{figure}
 \begin{center}
  \includegraphics[scale=0.8]{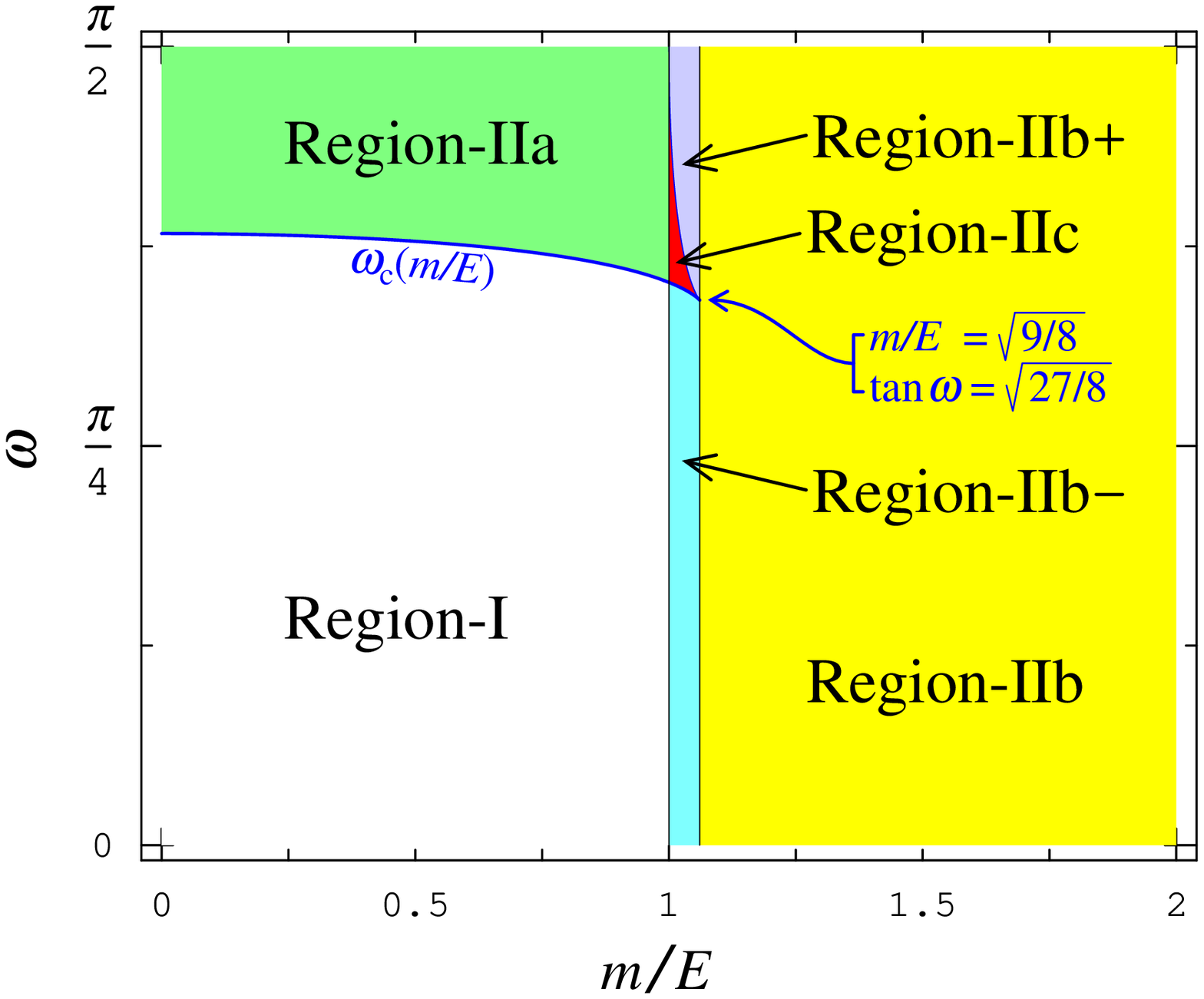}%
 \end{center}
 \caption{%
 Parameter space $(m/E, \omega)$ of the effective potential $W_\eff$
 for the ballistic particle.
 The space is classified into six regions;
 Region-I, Region-IIa,
 Region-IIb, Region-IIb$+$, Region-IIb$-$ and
 Region-IIc.
 The typical forms of the effective potential corresponding
 to the regions are shown in \fig{PotentialForms.eps}.
 The radiated particle runs away from the horizon in the Region-I.
 In the other regions (Region-II's),
 any radiated particle returns into the horizon.
 The Region-I is bounded on both
 the curve $\omega_c(m/E)$ in \EQ{omega_c} by the Region-IIa and
 the line $m/E=1$ by the Region-IIb$-$.
 We note $\tan\omega_c(0) = \sqrt{27/4}$ and $\tan\omega_c(1) = 2$.
 The Region-IIc is bounded
 on the curve $\omega_{c\dual}(m/E)$ in \EQ{omega_c_dual} by the Region-IIb$+$,
 on the curve $\omega_c(m/E)$ by the Region-IIb$-$
 and
 on the line $m/E=1$ by the Region-IIa.
 The Region-IIb is bounded on the line $m/E = \sqrt{9/8}$.
 \label{PhaseDiagram.eps}%
 }%
\end{figure}

\begin{figure}
 \begin{center}
  \includegraphics[scale=0.75]{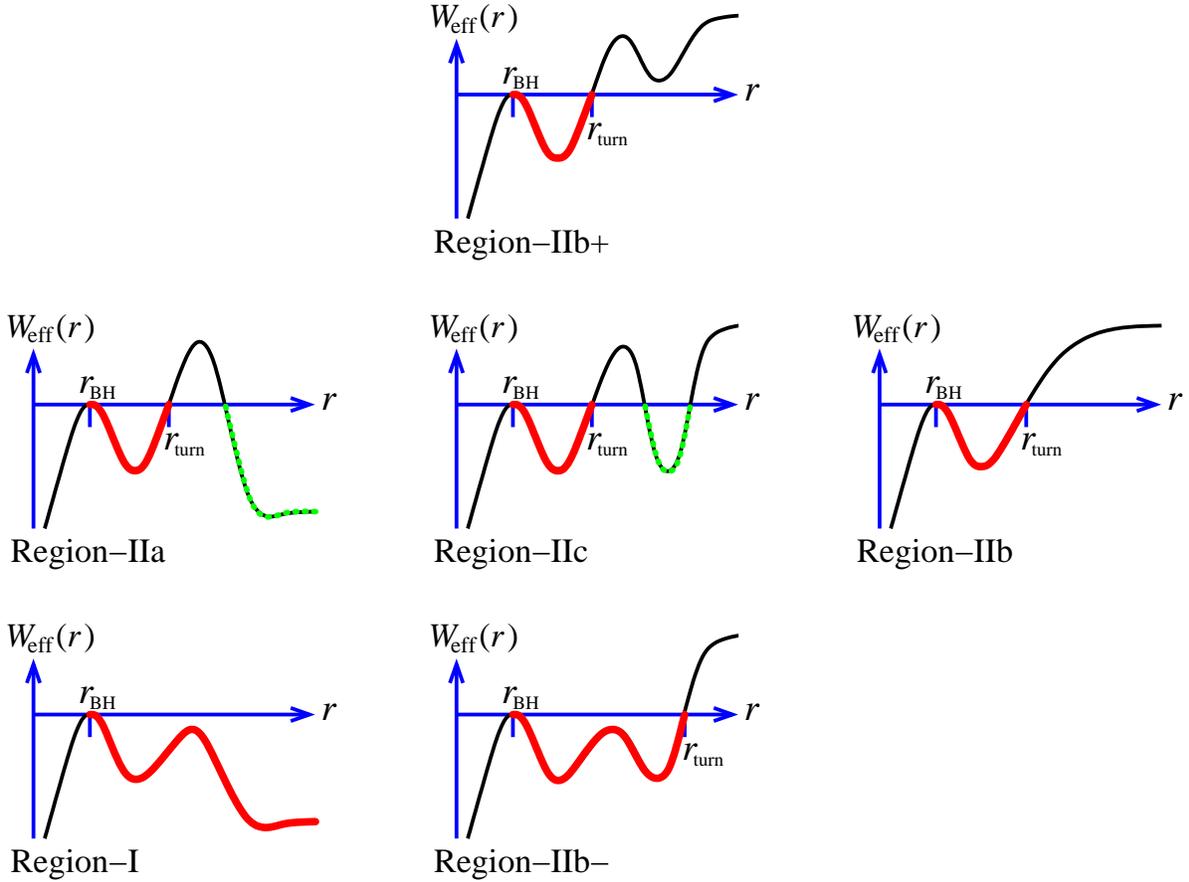}%
 \end{center}
 \caption{%
 Typical forms of the effective potential $W_\eff(r)$
 in \EQ{Particle-Effective-Potential-1}
 for the six parameter regions.
 The figures are arranged for corresponding to
 the position of each region
 in the parameter space in \fig{PhaseDiagram.eps}.
 The curve is the function-form of
 the effective potential $W_\eff(r)$ for the ballistic particle
 and the motion of the particle is allowed for $W_\eff(r) \leq 0$.
 The thick curve means the trajectory of particle-motion
 which is connected to the horizon.
 The motion of Hawking-radiated particle is described by the thick curves.
 The dotted curves mean the particle-trajectories disconnected to the horizon,
 i.e., the gravitational bending in the Region-IIa and
 the revolution-orbit in the Region-IIc,
 and is not relevant to our subject.
 In the Region-I the radiated particle runs away from the horizon
 into the infinite distance.
 In the Region-II's the radiated particle turns its $r$-direction
 at the radius $r = r_\turn$
 and returns into the horizon  $r = r_\BH$.
 \label{PotentialForms.eps}%
 }%
\end{figure}

According to the function-form of the effective potential
\EQ{Particle-Effective-Potential-1},
the parameter space $(m/E, \omega)$ is divided into the six regions;
the Region-I, the Region-IIa,
the Region-IIb, the Region-IIb$+$, the Region-IIb$-$
and the Region-IIc.
The division of the parameter space is shown in \fig{PhaseDiagram.eps}.
On each region
the effective potential $W_\eff$ has different forms
displayed in \fig{PotentialForms.eps}.
In the Region-I the particle radiated from the horizon
runs away from the horizon to the infinite distance
because the effective potential $W_\eff$ never cross the zero level
except for the horizon $r=r_\BH$.
In the Region-II's the radiated particle always 
turns its $r$-direction at each point satisfying $W_\eff(r) = 0$
and returns into the horizon $r=r_\BH$.
The trajectory of the radiated particle is connected with the horizon
and is described by each thick curve in \fig{PotentialForms.eps}.
In the Region-IIa and in the Region-IIc
there is a trajectory disconnected from the horizon
(the dotted curve in \fig{PotentialForms.eps}).
The trajectory disconnected from the horizon in the Region-IIc
means the revolution-orbit around the black hole
and that in the Region-IIa
means the gravitational-scattering-trajectory by the black hole.
Such trajectories disconnected from the horizon
are not relevant to our subject.

\begin{figure}
 \begin{center}
  \includegraphics[scale=1.0]{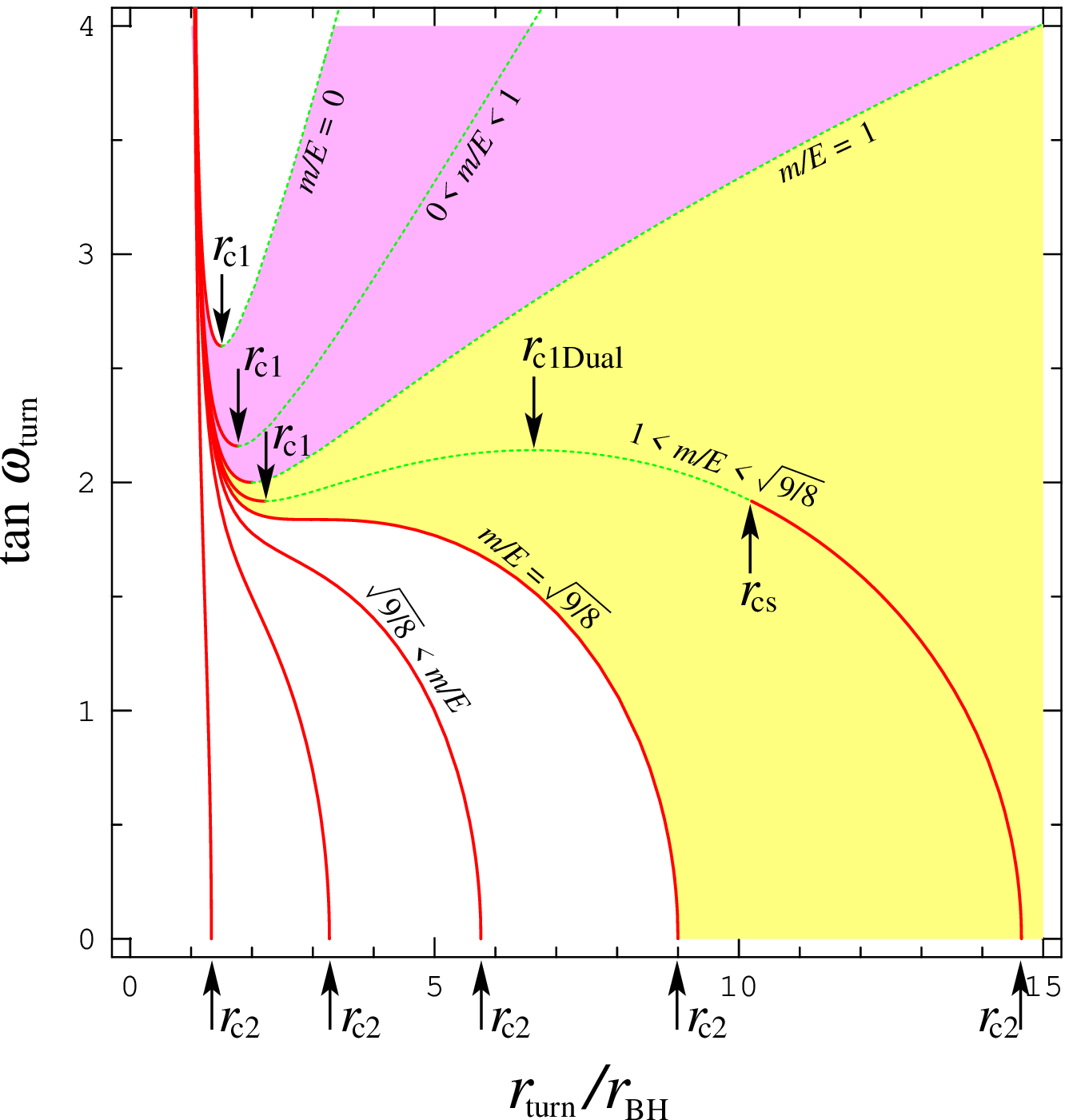}%
 \end{center}
 \caption{%
 The function-forms of $\tan\omega_\turn(r_\turn, m/E)$
 defined in \EQ{define-tan-omega} for various values of $m/E > 0$.
 The forms of the curve are classified into three classes
 $0 \leq m/E \leq 1$, $1 < m/E < \sqrt{9/8}$ and $\sqrt{9/8} \leq m/E$.
 The curve gives us the relation between the turning radius $r_\turn$
 and the radiation-zenith-angle $\tan\omega = \tan\omega_\turn(r_\turn, m/E)$.
 For given $\omega$,
 the solution $r$ nearest the horizon
 becomes the true turning radius $r_\turn$.
 Then solid curve is the turning radius $r_\turn$ and
 the dotted curve means a fake.
 For $0 \leq m/E < \sqrt{9/8}$,
 the solid curve is bounded on $r_\turn = r_\ci$
 which is the first maximum turning radius
 whose value is given in \EQ{r_ci}.
 On $\omega_\turn=0$,
 the solid curve is also bounded on $r_\turn = r_\cii$
 which is the second maximum turning radius \EQ{r_cii}.
 For $1 < m/E < \sqrt{9/8}$,
 we have another bound on $r_\turn = r_{\cs}$
 which is the splitting radius \EQ{splitting-radius}.
 \label{TanOmega.eps}%
 }%
\end{figure}

In the Region-II's
the relation amang the turning point $r_\turn$,
the radiation angle $\omega$ and the ratio $m/E$
is given by $W_\eff(r_\turn) = 0$ and it can be rewritten as
\begin{eqnarray}
 \tan^2\omega &=& \tan^2\omega_\turn\left(r_\turn, \frac{m}{E}\right),
 \label{omega-turn}
\end{eqnarray}
where we have defined a function
\begin{eqnarray}
 \tan^2\omega_\turn\left(r_\turn, \frac{m}{E}\right)
  &:=&
  \left(\frac{r_\turn}{r_\BH}\right)^2
  \left[\frac{1}{F(r_\turn)} - \left(\frac{m}{E}\right)^2  \right].
  \label{define-tan-omega}
\end{eqnarray}
We display the form of the function $\tan\omega_\turn$
in \fig{TanOmega.eps}.
When $0 \leq m/E < 1$,
the relation \EQ{omega-turn} is valid
for $r_\BH < r_\turn < r_\ci(m/E)$ as the Region-IIa.
When $1 \leq m/E < \sqrt{9/8}$,
the relation \EQ{omega-turn}
is valid for $r_\BH < r_\turn < r_\ci(m/E)$
as the Region-IIc and the Region-IIb$+$
and for $r_\cs < r_\turn < r_\cii(m/E)$ as the Region-IIb$-$.
The relation \EQ{omega-turn} is not valid for $r_\ci < r_\turn < r_\cs(m/E)$
because the relation indicates the turning radius of the revolution-orbit.
We note that any particle with $1 \leq m/E < \sqrt{9/8}$ does
not turn at the radius $r_\ci < r < r_\cs(m/E)$.
When $\sqrt{9/8} \leq m/E$, the relation
is valid for $r_\BH < r_\turn < r_\cii(m/E)$ as the Region-IIb.
We have defined the first maximum turning radius
\begin{eqnarray}
 r_\ci\left(\frac{m}{E}\right)
  &:=&
  \frac{r_\BH}{1-\left(\frac{m}{E}\right)^2}
  \left[
     \frac{3}{4} 
   - \left(\frac{m}{E}\right)^2
   + \frac{3}{4} \sqrt{1 - \frac{8}{9} \left(\frac{m}{E}\right)^2} \;
  \right]
  \label{r_ci}
\end{eqnarray}
for $0 < m/E < \sqrt{9/8}$ and its dual radius
\begin{eqnarray}
 r_{\ci\dual}\left(\frac{m}{E}\right)
  &:=&
  \frac{r_\BH}{1-\left(\frac{m}{E}\right)^2}
  \left[
     \frac{3}{4} 
   - \left(\frac{m}{E}\right)^2
   - \frac{3}{4} \sqrt{1 - \frac{8}{9} \left(\frac{m}{E}\right)^2} \;
  \right]
  \label{r_cidual}
\end{eqnarray}
for $1 < m/E < \sqrt{9/8}$.
These are given by the solutions of the quadratic equation
\begin{eqnarray}
 \left.\frac{\partial}{\partial r_\turn} \tan^2\omega_\turn
 \right|_{r_\turn = r_{\rm c}}  &=& 0.
\end{eqnarray}
We have also defined the second maximum turning radius
\begin{eqnarray}
 r_{\cii}\left(\frac{m}{E}\right)
  &:=&
  \frac{r_\BH \left(\frac{m}{E}\right)^2}{\left(\frac{m}{E}\right)^2-1}
  \label{r_cii}
\end{eqnarray}
for $1 < m/E$
as a solution of $\tan^2\omega(r_\cii, m/E) = 0$
and have defined the splitting radius as
\begin{eqnarray}
 r_\cs
  &:=&
  \frac{r_\BH}{\left(\frac{m}{E}\right)^2-1}
  \left[
     \frac{3}{2} 
   - \left(\frac{m}{E}\right)^2
   + \frac{3}{2} \sqrt{1 - \frac{8}{9} \left(\frac{m}{E}\right)^2} \;
  \right]
  \label{splitting-radius}
\end{eqnarray}
for $1 < m/E < \sqrt{9/8}$
as another solution of
$\tan^2\omega_\turn(r_\cs) = \tan^2\omega_\turn(r_\ci)$.
The critical curve $\omega = \omega_c(m/E)$
which separates the Region-I from the Region-IIa
in \fig{PhaseDiagram.eps} is given by
\begin{eqnarray}
 \tan^2\omega_c\left(\frac{m}{E}\right)
  &:=& \tan^2\omega_\turn\left(r_\ci, \frac{m}{E}\right) \nonumber\\
  &=& \frac{7}{2} - \left(\frac{m}{E}\right)^2 +
      \frac{1}{8}
      \frac{-1 + 27
      	    \left[1 - \frac{8}{9}\left(\frac{m}{E}\right)^2\right]^{3/2}}
      {1 - \left(\frac{m}{E}\right)^2}
  \label{omega_c}
\end{eqnarray}
where the right hand side is the function defined in \EQ{define-tan-omega}.
This critical curve also indicates the boundary
which separates the Region-IIc from the Region-IIb$-$.
This critical curve is defined for $0 \leq m/E \leq \sqrt{9/8}$.
The other boundary which separates the Region-IIc from the Region-IIb$+$
is given by
\begin{eqnarray}
 \tan^2\omega_{c\dual}\left(\frac{m}{E}\right)
  &:=& \tan^2\omega\left(r_{\ci\dual}, \frac{m}{E}\right) \nonumber\\
  &=& \frac{7}{2} - \left(\frac{m}{E}\right)^2 +
      \frac{1}{8}
      \frac{-1 - 27
      	    \left[1 - \frac{8}{9}\left(\frac{m}{E}\right)^2\right]^{3/2}}
      {1 - \left(\frac{m}{E}\right)^2}
  \label{omega_c_dual}
\end{eqnarray}
which is defined for $1 < m/E \leq \sqrt{9/8}$.

\subsection{Density Distribution as Functional of Flux}

We consider the density-distribution $\N_f(E,r)$.
The density-distribution
is a functional of the flux $f(E,\omega)$ on the horizon
and is also a function of both the parameter $m_f/E$ and the position $r$,
then we rewrite the density-distribution
as $\N(m/E,r; f)$ in this subsection.
The particle-density at the position $r$ is the
superposition of the density of the particles
which can reach the position $r$.
When we put values of both the radiation angle $\omega$
and the parameter $m/E$,
we find whether the particle can reach the position $r$ or not.
A contribution to the density by a particle is inversely proportional to
its velocity $\dot{y}^r$.
Therefore the particle-number-density is calculated by integrating out
the product of the flux $f(E,\omega)$ on the horizon,
the geometrical factor $\left(r_\BH/r\right)^2$
and the inverse velocity $1/\dot{y}^r$
with the integral variable $\omega$.
Here we define the function
\begin{eqnarray}
 \N_{\omega_1}^{\omega_2}(m/E,r; f)
  &:=&
  \left(\frac{r_\BH}{r}\right)^2
  \int_{\omega_1}^{\omega_2} d\omega \sin\omega
  \int_{0}^{2\pi} d\psi \;
  \frac{f(E,\omega)}{\dot{y}^r(r,\omega,m/E)}
  \label{Nomegaomega}
\end{eqnarray}
for convenience.
Because a part of the parameter space $(m/E, \omega)$
contributes to the density-distribution $\N(m/E,r; f)$ at the position $r$,
the integral interval $[\omega_1,\omega_2]$
for the density-distribution $\N(m/E,r; f)$
should be carefully chosen,
i.e.,
the integral interval $[\omega_1,\omega_2]$
depends on the position $r$ and on the parameter $m/E$.
The density-distribution is given by the sum of contributions
from the parameter regions:
\begin{eqnarray}
 \N\left(\frac{m}{E},r; f\right)
  &=&
    \N_{\I}	\left(\frac{m}{E},r;f\right)
  + \N_{\IIa}	\left(\frac{m}{E},r;f\right)
  + \N_{\IIb}	\left(\frac{m}{E},r;f\right)
  + \N_{\IIb\pm,{\rm c}}\left(\frac{m}{E},r;f\right).
  \label{Ntotal}
\end{eqnarray}
$\N_{\I}(m/E,r;f)$ is the density-distribution
contributed by the Region-I,
$\N_{\IIa}(m/E,r;f)$ is that by the Region-IIa,
$\N_{\IIb}(m/E,r;f)$ is that by the Region-IIb and
$\N_{\IIb\pm,{\rm c}}(m/E,r;f)$ is that by the combination of
the Region-IIb$-$, the Region-IIb$+$ and the Region-IIc.
These density-distributions are given by the function \EQ{Nomegaomega}
with a suitable integral interval $[\omega_1, \omega_2]$.
The density-distributions
$\N_{\I}(m/E,r;f)$, $\N_{\IIa}(m/E,r;f)$, $\N_{\IIb}(m/E,r;f)$
and $\N_{\IIb\pm,{\rm c}}(m/E,r;f)$
are considered as follows respectively.

\underline{1) The density-distribution contributed by the Region-I.}\\
%
%
This parameter-region is given by $0 \leq m/E < 1$
and $0 \leq \omega < \omega_c(m/E)$ (see \fig{PhaseDiagram.eps}).
The radiated particle in this region runs away from the horizon
$r=r_\BH$ to the infinite distance $r=\infty$
and never turns its $r$-direction (see \fig{PotentialForms.eps}).
%
%
Therefore the integral interval is given by $0 \leq \omega < \omega_c(m/E)$
and this interval does not depend on the position $r$.
The density-distribution becomes
\begin{eqnarray}
 \N_{\I} \left( \frac{m}{E},r;f \right) &=&
  \left\{
   \begin{array}{ll}
     \displaystyle
     \N_{0}^{\omega_c\left(\frac{m}{E}\right)}
     \left( \frac{m}{E},r;f \right)
     \quad
     & \left(0 \leq \frac{m}{E} < 1 \mbox{ and }
	r_\BH < r \leq \infty\right)\\[6mm]
     0
     & ({\rm others})
   \end{array}
  \right..
  \label{N_I-1}
\end{eqnarray}

\underline{2) The density-distribution contributed by the Region-IIa.}\\
%
%
This region is given by $0 \leq m/E < 1$
and $\omega_c(m/E) \leq \omega < \pi/2$ (see \fig{PhaseDiagram.eps}).
The radiated particle runs away from the horizon $r=r_\BH$,
turns its $r$-direction at radius $r=r_\turn$
which depends on the zenith-angle $\omega$
and returns into the horizon. (see \fig{PotentialForms.eps}).
Then the integral interval $[\omega_1,\omega_2]$
depends on the position $r$.
The particle within the parameter region
$\omega_c(m/E) \leq \omega \leq \omega_\turn(r,m/E)$
can reach the position $r$ and contributes to the density-distribution
$\N_{\IIa} (m/E,r; f)$
because
a particle with the zenith-angle $\omega = \omega_\turn(r,m/E)$
turns its $r$-direction at the position $r$.
On the other hand,
the particle within the parameter region
$\omega_\turn(r,m/E) < \omega < \pi/2$
cannot reach the position $r$ and
does not contribute to the density-distribution $\N_{\IIa} (m/E,r; f)$.
Therefore the integral interval should be
$\omega_c(m/E) \leq \omega \leq \omega_\turn(r,m/E)$
which is depending on the position $r$
and the density-distribution becomes
\begin{eqnarray}
 \N_{\IIa} \left(\frac{m}{E},r;f \right) &=&
  \left\{
   \begin{array}{ll}
     \displaystyle
     2 \times \N_{
     	\omega_c\left(\frac{m}{E}\right)}^{
	\omega_\turn\left(r,\frac{m}{E}\right)}
	\left(\frac{m}{E},r;f \right)
     \quad
     & \left(0 \leq \frac{m}{E} < 1 \mbox{ and }
        r_\BH < r \leq r_\ci\left(\frac{m}{E}\right)\right)\\[6mm]
     0
     & ({\rm others})
   \end{array}
  \right..
  \label{N_IIa-1}
\end{eqnarray}
We need the factor two in this form
because both the go and the return of the particle contribute
to the density-distribution.
Any particle in the Region-IIa does not contribute to
the density-distribution for $r > r_\ci\left(\frac{m}{E}\right)$
because the maximum turning radius in the Region-IIa
is given by $r_\ci\left(\frac{m}{E}\right)$
and particle does not turn on the radius $r > r_\ci\left(\frac{m}{E}\right)$.
Then the integral form in \EQ{N_IIa-1}
is restricted to $r_\BH < r < r_\ci\left(\frac{m}{E}\right)$.
For example
we obtain $\tan\omega_c = \sqrt{\frac{27}{4}}$ and $r_\ci = (3/2) r_\BH$
when we consider massless particle $m = 0$.
We obtain $\tan\omega_c = 2$ and $r_\ci = 2 r_\BH$
for the particle with $m/E = 1$.

\underline{3) The density-distribution contributed by the Region-IIb.}\\
%
%
This region is given by 
$\sqrt{9/8} \leq m/E$ and $0 \leq \omega < \pi/2$.
The particle whose zenith-angle satisfies
$0 \leq \omega \leq \omega_\turn\left(r, \frac{m}{E}\right)$
can reach the position $r$
because the radiated particle
with the zenith-angle $\omega = \omega_\turn\left(r, \frac{m}{E}\right)$
turns its $r$-direction at the position $r$.
The density at $r$ is given by the integration over the radiation angle
$0 \leq \omega \leq \omega_\turn\left(r, \frac{m}{E}\right)$.
We obtain
\begin{eqnarray}
 \N_{\IIb} \left(\frac{m}{E},r;f \right) &=&
  \left\{
   \begin{array}{ll}
     \displaystyle
     2 \times \N_{0}^{\omega_\turn\left(r, \frac{m}{E}\right)}
     \left(\frac{m}{E},r;f\right)
     \quad
     & \left(
	\sqrt{\frac{9}{8}} \leq \frac{m}{E} \mbox{ and }
        r_\BH < r \leq r_\cii\left(\frac{m}{E}\right)
       \right)\\[6mm]
     0
     & ({\rm others})
   \end{array}
  \right..
  \label{N_IIb-1}
\end{eqnarray}
The integral form in \EQ{N_IIb-1} is restricted to
$r_\BH < r \leq r_\cii\left(\frac{m}{E}\right)$
because the maximum turning radius in this region is given by
$r_\cii\left(\frac{m}{E}\right)$ rather than $r_\ci\left(\frac{m}{E}\right)$
(see the curves with $\sqrt{9/8} \leq m/E$ in \fig{TanOmega.eps}).
For example,
we have $r_\cii = 9 r_\BH$ for the particle
with $m/E = \sqrt{9/8}$ and
we have $r_\cii \rightarrow 1$ for the particle
with the limit $m/E \rightarrow \infty$.

\underline{4) The density-distribution contributed
by the Region-IIb$\pm$ and the Region-IIc.}\\
The parameter-region is simply given by
$1 \leq m/E < \sqrt{9/8}$ and $0 \leq \omega < \pi/2$,
however, the analysis becomes a little complicated
and more careful treatment is required
because the region is composed of three Regions (IIb$+$, IIb$-$ and IIc)
due to the revolution-orbit in the Region-IIc,
which is irrelevant to our subject.
%
%
Each particle radiated from the horizon with zenith-angle $\omega$
turns its $r$-direction at the position $r_\turn$
and returns into the horizon.
The relation between $\omega$ and $r_\turn$
is given by the equation \EQ{omega-turn} (also see \fig{TanOmega.eps}).
In the region-IIb$-$
($\omega \leq \omega_c(\frac{m}{E}) := \omega_\turn(r_\ci, \frac{m}{E})$) and
in the region-IIb$+$
($\omega > \omega_{c\dual}(\frac{m}{E}) := \omega_\turn(r_{\ci\dual}, \frac{m}{E})$),
we have a unique solution $r_\turn$
of the equation \EQ{omega-turn} for the fixed $\omega$.
On the other hand, in the region-IIc
($\omega_c < \omega \leq \omega_\turn(r_{\ci\dual})$)
there are three solutions of the equation \EQ{omega-turn}
(see \fig{PotentialForms.eps} and \fig{TanOmega.eps})
because
the revolution-orbit around the black hole is also allowed.
The solution nearest to the horizon
should be selected as $r_\turn$
because
the revolution-orbit which is disconnected with the horizon
is not relevant to the Hawking radiation
and
the trajectory of the Hawking radiation is connected to the horizon.

When the zenith-angle $\omega$ becomes larger from zero to $\omega_c(m/E)$
in the considering region $1 \leq m/E < \sqrt{9/8}$,
the turning radius $r_\turn$ becomes smaller.
When $\omega$ equals to $\omega_c(m/E)$,
the turning radius $r_\turn$ discontinuously jumps
from the radius $r_\cs$ to the smaller radius $r_\ci$
(see the curve with $1 < m/E < \sqrt{9/8}$ in \fig{TanOmega.eps})
because the parameter region is changed
from the Region-IIb$-$ into Region-IIc.
At the moment the trajectory is separated into the two disconnected parts
and one of the trajectories becomes the revolution-orbit which is irrelevant
(see the difference between $W_\eff(r)$ in the Region-IIb$-$
and that in Region-IIc in \fig{PotentialForms.eps}).
When $\omega$ becomes large from $\omega_c(m/E)$ to $\pi/2$,
the turning radius $r_\turn$ becomes small continuously
and approaches the horizon radius $r_\BH$.
Noting special happens in the transition
from the Region-IIc into the Region-IIb$+$
except for the disappearance of the revolution-orbit
which is not relevant.

To consider the density-distribution at the position $r$,
we should consider which particle can reach the position $r$.
When $r_\cs(\frac{m}{E}) \leq r \leq r_\cii(\frac{m}{E})$,
the particle belongs to the Region-IIb$-$.
The particle radiated
with the zenith-angle $\omega = \omega_\turn\left(r, \frac{m}{E}\right)$
turns its $r$-direction at the position $r$,
then
the particle with
$0 \leq \omega \leq \omega_\turn\left(r, \frac{m}{E}\right)$
can reach the position $r$.
The density at $r$ is given by the integration over the radiation angle
$0 \leq \omega \leq \omega_\turn\left(r, \frac{m}{E}\right)$.
When $r_\ci(\frac{m}{E}) < r < r_\cs(\frac{m}{E})$,
the particle still belongs to the Region-IIb$-$.
The radiated particle does not turn in the region
$r_\ci(\frac{m}{E}) < r < r_\cs(\frac{m}{E})$
and always turns at $r=r_\cs(\frac{m}{E})$.
Then the integral-interval should be
$0 \leq \omega \leq \omega_\turn\left(r_\cs, \frac{m}{E}\right)$.
The integral-interval can be rewritten as
$0 \leq \omega \leq \omega_c(\frac{m}{E})$
by using the definition of $\omega_c(\frac{m}{E})$ in \EQ{omega_c}.
When $r_\BH < r \leq r_\ci(\frac{m}{E})$,
the particle which belongs to the Region-IIc or the Region-IIb$+$
begins to contribute the density-distribution.
The situation is essentially same to the Region-IIb$-$
for $r_\cs(\frac{m}{E}) \leq r \leq r_\cii(\frac{m}{E})$
and the integral interval is also given by
$0 \leq \omega \leq \omega_\turn\left(r, \frac{m}{E}\right)$.
The density-distribution for $1 \leq m/E < \sqrt{9/8}$
is summarized as
\begin{eqnarray}
 \N_{\IIb\pm,{\rm c}} \left(\frac{m}{E},r;f \right) &=&
  \left\{
   \begin{array}{ll}
     0 & \left(
	 r_\cii\left(\frac{m}{E}\right) < r \leq \infty
	\right)\\[6mm]
     2 \times \N_{0}^{ \omega_\turn\left(r, \frac{m}{E}\right) }
      \left(\frac{m}{E},r;f\right)
      \quad
      & \left(
	 r_\cs\left(\frac{m}{E}\right) \leq r \leq
	 r_\cii\left(\frac{m}{E}\right)
	\right)\\[6mm]
     2 \times \N_{0}^{\omega_c\left(\frac{m}{E}\right)}
      \left(\frac{m}{E},r;f\right)
      \quad
      & \left(
	 r_\ci\left(\frac{m}{E}\right) < r <
	 r_\cs\left(\frac{m}{E}\right)
	\right)\\[6mm]
     2 \times \N_{0}^{ \omega_\turn\left(r, \frac{m}{E}\right) }
      \left(\frac{m}{E},r;f\right)
      \quad
      & \left(
	 r_\BH < r \leq
	 r_\ci\left(\frac{m}{E}\right)
	\right)
   \end{array}
  \right..
  \label{N_IIc-1}
\end{eqnarray}
The resultant form \EQ{N_IIc-1} has a little complicated conditions for $r$,
however, the form is a continuous function of $r$
because the integral interval
continuously changes as $r$ changes.

\section{Radiation Angle Distribution}\label{angle.sec}

In this section we consider the particle-flux on the horizon
$f(E,\omega)$ in \EQ{Diff-Flux},
which is required to evaluate
the particle density-distribution $\N(m/E,r;f)$ in \EQ{Ntotal}.
%
%

As a consequence of the assumption-(i)
which is mentioned in the previous section
and of the spherical symmetry,
the observer at the infinite distance finds a disk image
as a detection of the Hawking-radiated particles.
In \fig{ImpactParam.eps}
the correspondence of a particle-geodesic
to a point in the image of the radiation on the screen
at the infinite distance
is schematically shown.
The observed image is equal to the image on the screen.
The screen is parameterized by the polar coordinate $(D,\chi)$,
where the parameter $D$ is distance from the center of the screen.
The parameter $D$ can be regard the impact parameter
of an absorption-process into the black hole
because the absorption of the particle
is regarded as the inverse process of the Hawking radiation.
A relation
\begin{eqnarray}
 \frac{L}{E} &=& v_\infty D,
\end{eqnarray}
is derived from conservation of the angular momentum of the particle,
where velocity of the particle at the infinite distance
\begin{eqnarray}
 v_\infty\left(\frac{m}{E}\right) &:=& \sqrt{1 - \left(\frac{m}{E}\right)^2}
  \label{v_infty}
\end{eqnarray}
is defined.
The particle which reaches the infinite distance satisfies a condition $E>m$.
By combining \EQ{v_infty} with \EQ{tanOmegaAndLE}
we obtain the relation between
the radiation-zenith-angle $\omega$ and the impact parameter $D$ as
\begin{eqnarray}
 D\left(\frac{m}{E}, \: \omega\right)
 &=& r_\BH \frac{\tan\omega}{v_\infty}.
 \label{D-TanOmega}
\end{eqnarray}
In order for a radiated particle to reach the infinite distance,
the radiation-zenith-angle should satisfy
$0 \leq \tan\omega < \tan\omega_c(\frac{m}{E})$.
This restriction for the radiation-zenith-angle $\omega$
is corresponding to
the restriction for the impact parameter $0 \leq D < D_\crit(\frac{m}{E})$.
By using the relation between $\omega$ and $D$ in \EQ{D-TanOmega},
we obtain the radius of the apparent disk-image on the screen as
\begin{eqnarray}
 D_\crit\left(\frac{m}{E}\right)
 &:=& r_\BH \frac{\tan\omega_c}{v_\infty}.
\end{eqnarray}
Therefore the impact parameter $D$, namely the position of the image,
is one-to-one corresponding to the radiation-zenith-angle $\omega$
if $D < D_\crit(m/E)$ and $E > m$ are satisfied.

The apparent disk area
\begin{eqnarray}
 \sigma_\BH\left(\frac{m}{E}\right)
  &:=& \pi D_\crit^2\left(\frac{m}{E}\right)
  \label{abs-cross-section}
\end{eqnarray}
means the absorption-cross-section of the black hole for the particle
in the ballistic picture (eikonal limit).
The area of the disk is depending on $m/E$.
For example
we have $D_\crit = \sqrt{\frac{27}{4}} r_\BH$ and
$\sigma_\BH = \frac{27}{4} \pi r_\BH^2$ for massless particles.

Because the Hawking radiation for the observer at the infinite distance
is regarded as an inverse process of the particle-absorption
into the horizon,
the differential luminosity of the Hawking radiation is given by
the product of the absorption-cross-section and
the thermal flux from the black body per unit area:
\begin{eqnarray}
 dL (T_\BH, m, E) &=&
  \left\{
   \begin{array}{ll}
    \displaystyle
    \sigma_\BH\left(\frac{m}{E}\right)
     \times \frac{v_\infty}{4} 
     \times \frac{g}{(2\pi)^3} f_{T_\BH}(E) \: 4\pi E^2 v_\infty \: dE
      & (E > m) \\[2mm]
    0 & (E < m)
   \end{array}
  \right.,
  \label{dL_f}
\end{eqnarray}
where
\begin{eqnarray}
 f_{T_\BH}(E) &:=& \frac{1}{e^{\frac{E}{T_\BH}} \pm 1}
\end{eqnarray}
is the Fermi-Dirac or the Bose-Einstein distribution-function
for the Hawking temperature $T_\BH$ and
$g$ is the degree of the freedom for the particle
on the temperature $T_\BH$.
The absorption-cross-section $\sigma_\BH\left(\frac{m}{E}\right)$
in the luminosity $dL$ is depending on the particle-energy $E$.
This dependency is referred as ``the gray body factor''
of the Hawking radiation
\cite{Hawking:1975sw,Hawking:1974rv,Maldacena:1996ix}.
We have a relation between
the flux on the horizon $f(E,\omega)$ in \EQ{Diff-Flux} and
the differential luminosity \EQ{dL_f}:
\begin{eqnarray}
 {4 \pi r_\BH^2} \int d\F
  \;=\;
  {4 \pi r_\BH^2}
  \int_0^{\omega_c} d\omega \sin\omega \int_0^{2\pi} d\chi \:
  f(E,\omega) dE
  &=&
  dL(T_\BH,m,E).
  \label{f-L-Consistency}
\end{eqnarray}
This relation gives us a restriction for the flux $f(E,\omega)$.

To determine the flux $f(E,\omega)$, we assume 
\begin{enumerate}
 \item[(ii)] the brightness of the disk-image for each particle-energy 
	     appears to be uniform
	     over the disk-image for the observer at the infinite distance
	     and
	     the energy-spectrum of the disk-image is given by
	     the perfect black body radiation with the Hawking temperature.
\end{enumerate}
By the above assumption and the relation \EQ{D-TanOmega}
we have the relation between
the differential brightness of the disk \EQ{dL_f}
and the flux on the horizon as
\begin{eqnarray}
  & &
  \frac{v_\infty}{4} \times
  \frac{g_{f}}{(2\pi)^3} f_{T_\BH}(E) \: 4\pi E^2 v_\infty dE \times
  (dD D d\chi) \nonumber\\
  &=&
  4\pi r_\BH^2
  \times
  \left[
   \frac{1}{4\pi} \frac{1}{\cos^3\omega}
   \times
   \frac{1}{4}  
   \frac{g_{f}}{(2\pi)^3} f_{T_\BH}(E) \: 4\pi E^2 dE
  \right]
  \times
  (d\omega \sin\omega d\chi).
  \label{flux-luminosity-correspondence}
\end{eqnarray}
By using the spherical symmetry of the Hawking radiation,
$(d\omega\sin\omega d\chi)$
in the right hand side of \EQ{flux-luminosity-correspondence}
can be rewritten to $(d\omega\sin\omega d\psi)$\footnote{
$d\chi$ is equal to $d\phi$
due to the configuration in \fig{ImpactParam.eps}.
$d\phi$ can be regard as $d\psi$
because both the space-time and the Hawking-radiation
are spherical symmetric then $\phi$-axis is exchangeable for $\psi$-axis.}.
By comparing with the definition of the particle-flux in \EQ{Diff-Flux},
we obtain the particle-flux on the horizon per unit area
\begin{eqnarray}
 f(E,\omega)
  &=&
   \frac{1}{16\pi} \frac{1}{\cos^3\omega}
   \times
   \frac{g}{(2\pi)^3} f_{T_\BH}(E) \: 4\pi E^2.
   \label{result-flux}
\end{eqnarray}
The resultant flux \EQ{result-flux}
satisfies the restriction in \EQ{f-L-Consistency}.

We should note that the resultant flux \EQ{result-flux}
is defined for the particles which belong to the Region-I
because the particles in Region-II's cannot reach the observer.
Namely, the flux \EQ{result-flux} is only defined for
high elevation-angle\footnote{
The elevation-angle of the radiation is defined as $(\pi/2 - \omega)$.
High elevation-angle means small zenith-angle $\omega \sim 0$ and
low elevation-angle means large zenith-angle $\omega \sim \pi/2$.}
$0 \leq \omega < \omega_c$
and high particle-energy $E > m_f$.
We need the flux $f(E,\omega)$ defined for all elevation-angle
$0 \leq \omega < \pi/2$ and for all particle-energy $0 \leq E \leq \infty$
to calculate the particle number density $\N(m/E,r;f)$.
To solve the problem
we perform a analytic continuation of the flux, namely,
we extend the domain of the flux \EQ{result-flux}
to all elevation-angle $0 \leq \omega < \pi/2$
and all particle-energy $0 \leq E \leq \infty$
with keeping the function form of the flux \EQ{result-flux}.
The analytic continuation can be performed $C^\infty$-smoothly
because no singularity exists in the new domain of definition.
The resultant flux \EQ{result-flux} defined for $0 \leq \omega < \pi/2$
diverges on $\omega = \pi/2$.

The radiation-flux on the normal black body with temperature $T_\BH$
is given by
\begin{eqnarray}
 f(E,\omega)
  &=&
   \frac{\cos\omega}{4\pi}
   v^2
   \times
   \frac{g}{(2\pi)^3} f_{T_\BH}(E) \: 4\pi E^2.
   \label{black-body-flux}
\end{eqnarray}
Our resultant flux on the horizon \EQ{result-flux} is
different from the flux \EQ{black-body-flux}.
Especially the dependence on the radiation-zenith-angle $\omega$
is quite different.
We find the low-elevation-angle-dominance of the flux on the horizon.
The dependence on the particle-velocity is also different.

\section{Density of the Ballistic Particles II}\label{density2.sec}

We concretely calculate the particle-density-distribution \EQ{Ntotal}
by using the resultant flux on the horizon \EQ{result-flux}.
We prepare several functions
to calculate the density-distribution \EQ{Ntotal}.
By substituting the flux $f(E,\omega)$ resulted in \EQ{result-flux}
into the functional $\N_{\omega_1}^{\omega_2}(m/E,r; f)$ in \EQ{Nomegaomega},
we obtain
\begin{eqnarray}
  \N_{\omega_1}^{\omega_2}(m/E,r)
  &=&
   \frac{1}{16\pi} \frac{g}{(2\pi)^3} f_{T_\BH}(E) \: 4\pi E^2
   \left(\frac{r_\BH}{r}\right)^2  
   \int_{\omega_1}^{\omega_2} d\omega
   \int_{0}^{2\pi} d\chi \;
   \frac{\sin\omega}{\cos^3\omega}
   \frac{1}{\dot{y}^r(r,\omega,m/E)} \nonumber\\
  &=&
   \frac{1}{8} \frac{g}{(2\pi)^3} f_{T_\BH}(E) \: 4\pi E^2
   \left(\frac{r_\BH}{r}\right)^2
   \frac{1}{F(r)}
   \tilde{G}_{\omega_1}^{\omega_2} (A,B),
   \label{Nomegaomega2}
\end{eqnarray}
where we have defined
\begin{eqnarray}
 \tilde{G}_{\omega_1}^{\omega_2}(A,B)
  &:=&
  \int_{\omega_1}^{\omega_2} d\omega
  \frac{\sin\omega}{\cos^3\omega}
  \frac{1}{\sqrt{B - A \tan^2\omega}}
  \label{G-Function}\\
 A &:=& \left(\frac{r_\BH}{r}\right)^2 F(r) \\
 B &:=& 1 - \left(\frac{m}{E}\right)^2 F(r).
\end{eqnarray}
The definite integration in \EQ{G-Function}
can be performed as
\begin{eqnarray}
 \tilde{G}_{\omega_1}^{\omega_2}(A,B)
  &:=&
  \tilde{G}_{\omega_1}(A,B) - \tilde{G}_{\omega_2}(A,B),
\end{eqnarray}
where
\begin{eqnarray}
 \tilde{G}_\omega &:=& \frac{1}{A}\sqrt{B - A \tan^2\omega}
  \label{primitive-function}
\end{eqnarray}
is a primitive function of the integrand.
For convenience of the following calculations
we show several evaluations of the primitive function \EQ{primitive-function}:
\begin{eqnarray}
 \tilde{G}_0
 &=&
 \frac{1}{F} \left(\frac{r}{r_\BH}\right)^2
  \sqrt{1 - \left(\frac{m}{E}\right)^2 F(r)}
  \label{tildeG_0}\\
 \tilde{G}_{\omega_c(\frac{m}{E})}
 &=&
 \frac{1}{F} \left(\frac{r}{r_\BH}\right)^2
  \frac{|r - r_\ci|}{r}
  \sqrt{ \left\{\left(\frac{m}{E}\right)^2 - 1 \right\}
         \frac{r_\cs - r}{r} }
  \label{tildeG_omega_c}\\
 \tilde{G}_{\omega_\turn(r,\frac{m}{E})}
 &=&
  0.
  \label{tildeG_omega_turn}
\end{eqnarray}

We rewrite the density-distribution $\N(E,r;f)$ in \EQ{Ntotal} as 
\begin{eqnarray}
 \N(E,r) &=& 
   \frac{1}{4}
   \frac{g}{(2\pi)^3} f_{T_\BH}(E) \: 4\pi E^2
   \; \times \; 
   \left(\frac{r_\BH}{r}\right)^2
   \frac{1}{F^2(r)} \;
   \frac{1}{2}
   G\left(\frac{m}{E}, r\right),
 \label{N_f-G_f}
\end{eqnarray}
where we have defined a dimensionless positive function $G$ as
\begin{eqnarray}
 G\left(\frac{m}{E},r\right)
  &=&
     G_{\I}	\left(\frac{m}{E},r\right)
   + G_{\IIa}	\left(\frac{m}{E},r\right)
   + G_{\IIb\pm,{\rm c}}\left(\frac{m}{E},r\right)
   + G_{\IIb}	\left(\frac{m}{E},r\right).
  \label{Gtotal}
\end{eqnarray}
The functions
$G_{\I}, G_{\IIa}, G_{\IIb\pm,{\rm c}}$ and $G_{\IIb}$
are corresponding to the partial density-distributions
$\N_{\I}, \N_{\IIa}, \N_{\IIb\pm,{\rm c}}$ and $\N_{\IIb}$
in \EQ{Ntotal}.
The relation between $G$'s and $\N$'s
is also given by \EQ{N_f-G_f}.
The factors before a symbol of the first product in \EQ{N_f-G_f}
correspond to the number-density of the particles
radiated from the normal black body with temperature $T_\BH$.
The factor $(r_\BH/r)^2$ is a geometrical factor
of spherical source of the radiation.
The factor $1/F^2(r)$ means a principal general relativistic effect
and plays important role near the horizon.
The factor $G/2$ means a residual correction
including the finite mass effects.
We can show that $G/2$ is regular function
and have finite value $G(r_\BH)/2 = 1$ on the horizon.
We can also show that $G$ is finite for all position $r$
except for $G(1, r \rightarrow \infty)$.
Therefore the density-distribution \EQ{N_f-G_f}
is governed by the factor $1/F^2(r)$ near the horizon.
When $m<E$, the density-distribution \EQ{N_f-G_f}
is governed by the factor $(r_\BH/r)^2$ for distant region.

We concretely calculate the correction factors
$G_{\I}, G_{\IIa}, G_{\IIb\pm,{\rm c}}$ and $G_{\IIb}$
by using the relations \EQ{tildeG_0}, \EQ{tildeG_omega_c}
and \EQ{tildeG_omega_turn}.
From \EQ{N_I-1} we obtain
\begin{eqnarray}
 G_{\I} \left(\frac{m}{E},r\right)
  &=&
  \left(\frac{r}{r_\BH}\right)^2
  \left[
     \sqrt{1 - \left(\frac{m}{E}\right)^2 F}
   - \frac{|r - r_\ci|}{r}
     \sqrt{ \left\{\left(\frac{m}{E}\right)^2 - 1 \right\}
     \frac{r_\cs - r}{r} }
  \right]
  \label{G_I}
\end{eqnarray}
for $r_\BH \leq r \leq \infty$.
We find
\begin{eqnarray}
 \lim_{r\rightarrow\infty}
  G_{\I} \left(\frac{m}{E}, r\right)
  &=&
  \frac{\tan\omega_c^2(\frac{m}{E})}{v_\infty}
  \label{G_I-limit}
\end{eqnarray}
and it is consistent with the relation
\begin{eqnarray}
 \lim_{r\rightarrow\infty}
 \frac{dL}{4\pi r^2}
  &=& 
 \lim_{r\rightarrow\infty}
  v_\infty \N dE
\end{eqnarray}
between \EQ{dL_f} and \EQ{N_f-G_f}.
From \EQ{N_IIa-1} we obtain
\begin{eqnarray}
 G_{\IIa} \left(\frac{m}{E},r\right)
  &=&
  \left\{
   \begin{array}{ll}
     2 \times
     \left(\frac{r}{r_\BH}\right)^2
     \frac{|r - r_\ci|}{r}
     \sqrt{ \left\{\left(\frac{m}{E}\right)^2 - 1 \right\}
            \frac{r_\cs - r}{r} }
    & (r_\BH \leq r \leq r_\ci)\\
    0 & (r_\ci < r \leq \infty)
   \end{array}
  \right..
  \label{G_IIa}
\end{eqnarray}
By using the property of the absolute-value function in \EQ{G_I}
and \EQ{G_IIa},
we can summarize 
\begin{eqnarray}
 G_{\I + \IIa} \left(\frac{m}{E},r\right)
  &:=&
    G_{\I}	\left(\frac{m}{E},r\right)
  + G_{\IIa}	\left(\frac{m}{E},r\right)
  \\
  &=&
  \left(\frac{r}{r_\BH}\right)^2
  \left[
     \sqrt{1 - \left(\frac{m}{E}\right)^2 F}
   - \frac{r - r_\ci}{r}
     \sqrt{ \left\{\left(\frac{m}{E}\right)^2 - 1 \right\}
     \frac{r_\cs - r}{r} }
  \right].
  \label{G_I+IIa}
\end{eqnarray}
as the $G$-function for $0 \leq m/E < 1$ and $r_\BH \leq r \leq \infty$.
Both $G_{\I}$ and $G_{\IIa}$ as a function of $r$
are non-differentiable at $r=r_\ci$, however,
the function $G_{\I + \IIa}$ becomes differentiable at $r=r_\ci$.
The function $G_{\I + \IIa}$ is smooth
and monotonously increasing function for $r_\BH \leq r \leq \infty$.
The function has the lower bound $2$ for $r=r_\BH$ and
the upper bound \EQ{G_I-limit}.
The function $G_{\I + \IIa}$ is almost flat for $r \gg r_\BH$
except for $m/E = 1$,
then it can be approximated with
$G_{\I + \IIa}(m/E, r) \simeq G_{\I}(m/E, \infty)$ for $r \gg r_\BH$.

From \EQ{N_IIb-1} we obtain
\begin{eqnarray}
 G_{\IIb} \left(\frac{m}{E}, r\right)
  &=&
  \left\{
   \begin{array}{ll}
     2 \times
     \left(\frac{r}{r_\BH}\right)^2
     \sqrt{1 - \left(\frac{m}{E}\right)^2 F(r)}
    & \left( r_\BH \leq r \leq r_\cii\left(\frac{m}{E}\right) \right)\\
     0
    & \left( r_\cii\left(\frac{m}{E}\right) < r \leq \infty \right)
   \end{array}
  \right.
  \label{G_IIb}
\end{eqnarray}
This result has contained all contribution
from the parameter region $\sqrt{9/8} \leq m/E$.
This is a finite and continuous function of $r$.
This is also differentiable except for $r = r_\cii$.

From \EQ{N_IIc-1} we obtain
\begin{eqnarray}
 && G_{\IIb\pm,{\rm c}} \left(\frac{m}{E}, r\right) =
  2 \times
     \left(\frac{r}{r_\BH}\right)^2
  \nonumber\\
 &&
  \times
  \left\{
   \begin{array}{ll}
     \sqrt{1 - \left(\frac{m}{E}\right)^2 F(r)}
      & \left(
	  r_\BH \leq r \leq
	  r_\ci\left(\frac{m}{E}\right)
	\right)\\[6mm]
      \left[
         \sqrt{1 - \left(\frac{m}{E}\right)^2 F}
       - \frac{|r - r_\ci|}{r}
         \sqrt{ \left\{\left(\frac{m}{E}\right)^2 - 1 \right\}
	 	\frac{r_\cs - r}{r} }
      \right] \quad
      & \left(
	 r_\ci\left(\frac{m}{E}\right) < r <
	 r_\cs\left(\frac{m}{E}\right)
	\right)\\[6mm]
     \sqrt{1 - \left(\frac{m}{E}\right)^2 F(r)}
      & \left(
	  r_\cs\left(\frac{m}{E}\right) \leq r \leq
	  r_\cii\left(\frac{m}{E}\right)
	\right)\\[6mm]
     0
      & \left(
	 r_\cii\left(\frac{m}{E}\right) < r \leq \infty
	\right)
   \end{array}
  \right..
  \label{G_IIc-1}
\end{eqnarray}
for $1 < m/E < \sqrt{9/8}$.
This is a finite and continuous function of $r$, however,
is not differentiable at $r = r_\ci, r_\cs$ and $r_\cii$.

When $m/E = 1$, the function $G_\I$ and $G_{\IIb\pm,{\rm c}}$
diverge for the limit $r\rightarrow\infty$
because the particle velocity at the infinite distance $v_\infty$
becomes zero.
However this divergence does not crucially affect
the effective potential \EQ{Effective-Potential-3}
because
the integration by $E$ in \EQ{mu_eff} moderates the divergence
and the effective Higgs mass \EQ{mu_eff} becomes finite
for $r\rightarrow\infty$.

\section{Dynamical Formation of Spherical Domain Wall}\label{formation.sec}

In this section we concretely derive the effective potential for the
Higgs vev and consider the wall formation.
The effective potential \EQ{Effective-Potential-3}
is governed by the effective Higgs mass $\mu_\eff^2(r; \N)$
defined in \EQ{mu_eff}.

\subsection{Effective Higgs Mass around Black Hole}

\begin{figure}
 \begin{center}
  \begin{tabular}{c@{\hspace{6mm}}c}
     \hspace*{-5mm}
     \includegraphics[scale=0.49]{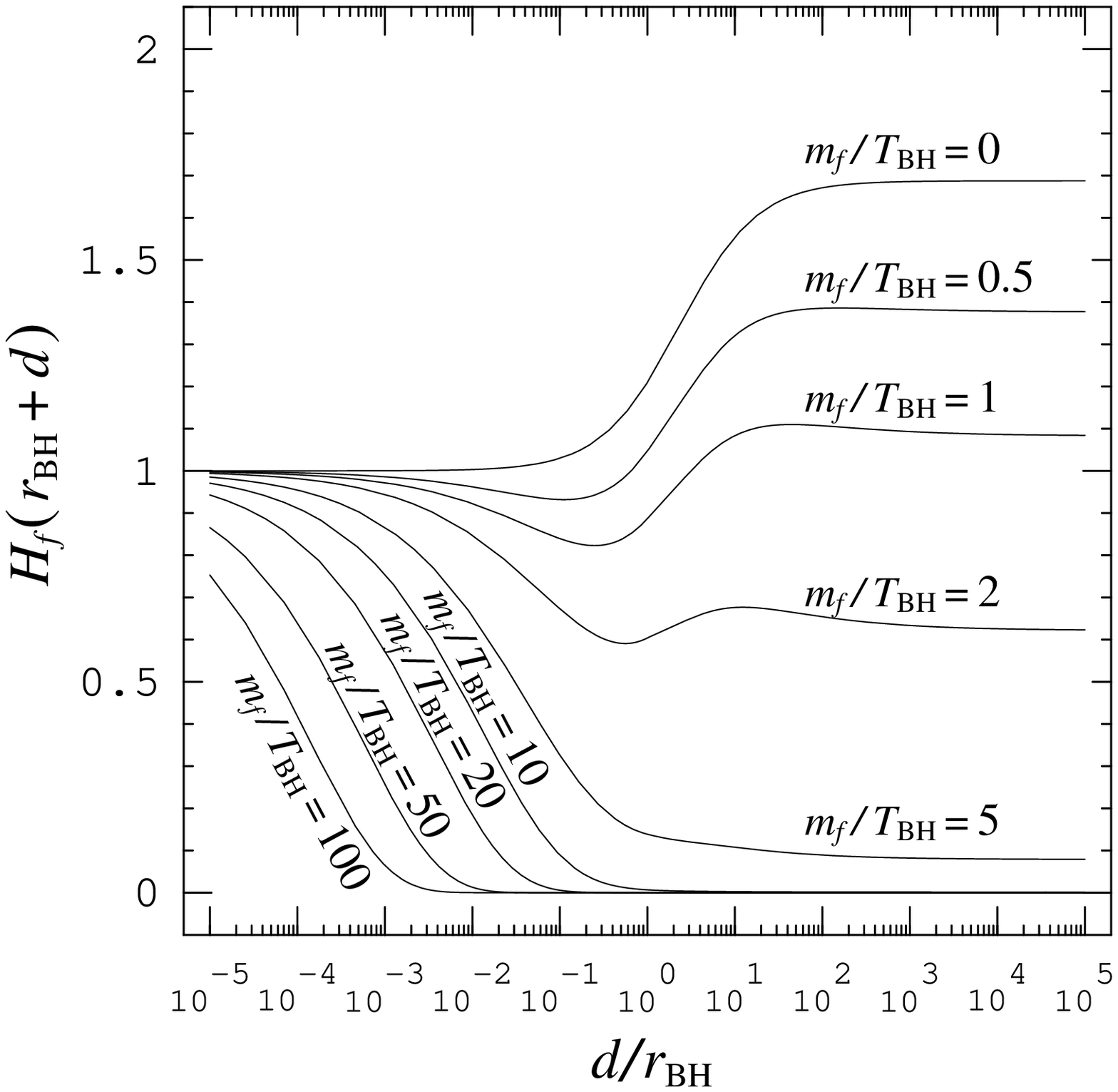}
   & \includegraphics[scale=0.49]{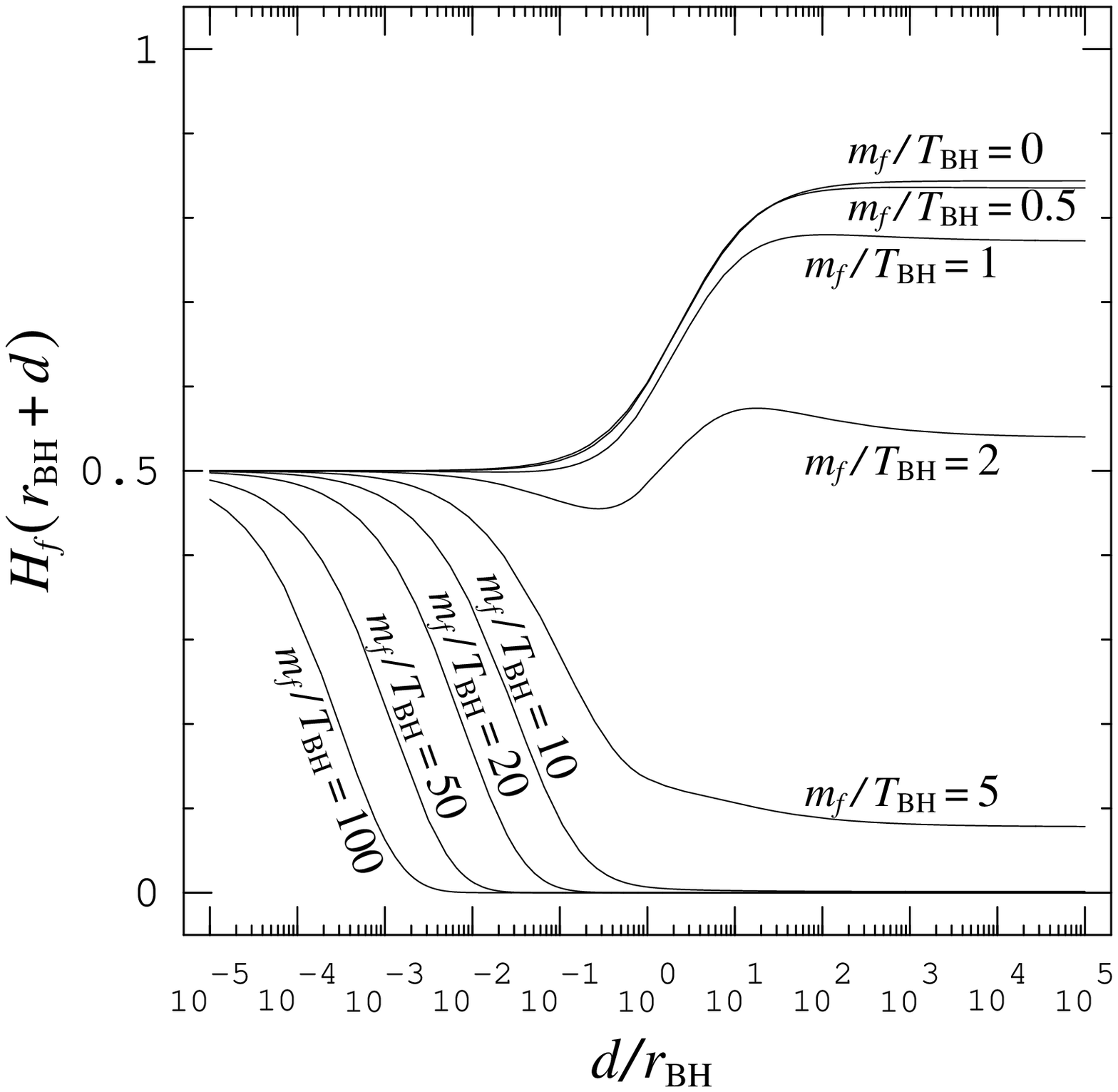}\\
     ({\bf a}) Boson
   & ({\bf b}) Fermion
  \end{tabular}
 \end{center}
 \caption{%
 Shapes of the form-function $H_f(\frac{m_f}{T_\BH},r)$
 defined in \EQ{H_f-1}
 for various $m_f/T_\BH$ and for ({\bf a}) bosons and for ({\bf b}) fermions.
 The horizontal axis $d/r_\BH$ is the distance from the horizon
 $d = r - r_\BH$,
 which is normalized by the Schwarzschild radius $r_\BH$.
 For any $m_f/T_\BH$, we have the near-horizon limits
 $H_f(r\rightarrow r_\BH) \rightarrow 1$   for bosons and
 $H_f(r\rightarrow r_\BH) \rightarrow 1/2$ for fermions.
 For massless particles, we have
 $H_f(r\rightarrow \infty) \rightarrow 27/16 =1.6875$ for bosons and
 $H_f(r\rightarrow \infty) \rightarrow 27/32$ for fermions.
 \label{Hf.eps}%
 }%
\end{figure}

By substituting the particle-density-distribution $\N(E,r)$
resulted in \EQ{N_f-G_f},
the effective Higgs mass $\mu_\eff^2(r; \N)$ in \EQ{mu_eff} becomes
\begin{eqnarray}
 \mu_\eff^2(r)
  &=& -\mu^2
  \;+\;
  \frac{1}{8\pi^2}
  \frac{1}{F(r)}
  \left(\frac{r_\BH}{r}\right)^2
  \sum_f g_f Y_f^2
  \int_0^\infty dE E
   f_{T_\BH}(E) \: 
   \frac{1}{2} 
   G\left(\frac{m_f}{E}, r\right).
  \label{mu_eff-2}
\end{eqnarray}
By defining a form function
\begin{eqnarray}
 H_f\left(\frac{m_f}{T_\BH},r\right)
  &=&
  \frac{6}{\pi^2}
  \int_0^\infty ds \; s f_1(s) \frac{1}{2}
  G \left(\frac{m_f}{T_\BH}\frac{1}{s}, r\right),
  \label{H_f-1}
\end{eqnarray}
we can rewrite
\begin{eqnarray}
 \mu_\eff^2(r)
  &=& -\mu^2
  \;+\;
  \frac{1}{48}
  \frac{1}{F(r)}
  \left(\frac{r_\BH}{r}\right)^2
  T_\BH^2
  \sum_f g_f Y_f^2
  H_f\left(\frac{m_f}{T_\BH},r\right).
  \label{mu_eff-3}
\end{eqnarray}
The shapes of the form function \EQ{H_f-1} are shown in \fig{Hf.eps}.
The form function \EQ{mu_eff-3} is defined for $r_\BH \leq r \leq \infty$.
The range of the form function \EQ{mu_eff-3} becomes
$0 \leq H_f(m/T_\BH,r) \leq 27/16$ for bosons
and $0 \leq H_f(m/T_\BH,r) \leq 27/32$ for fermions.

{\it Non General Relativistic Limit} ---
If we omit the general relativistic (GR) effects
as $F(r)=1$ and $\frac{1}{2} G_f = 1$,
we obtain $H_f = 1$ for any boson
and $H_f = \frac{1}{2}$ for any fermion.
Then the effective Higgs mass becomes
\begin{eqnarray}
 \mu_\eff^2(r)
  &=& -\mu^2
  \;+\;
  \frac{A^2}{r^2},
  \label{mu_eff-old}
\end{eqnarray}
where we have defined {\it the wall formation constant} as
\begin{eqnarray}
 A^2 &:=& \frac{1}{768\pi^2}\sum_f \tilde{g}_f Y_f^2
 \label{wall-formation-constant}
\end{eqnarray}
and an effective degree of the freedom as
\begin{eqnarray}
 \tilde{g}_f &:=&
  \left\{
   \begin{array}{ll}
    g_f & (f : {\rm boson})\\
    \frac{1}{2}g_f \quad & (f : {\rm fermion})
   \end{array}
  \right..
  \label{tilde-g_f}
\end{eqnarray}
The effective Higgs mass \EQ{mu_eff-old} reproduces
the result in our previous work \cite{BHBG3}
where the GR effects including the red-shift-effect near the horizon and
the gray-body-factor are omitted.
%

{\it High Temperature Limit} ---
When we consider the high temperature limit $T_\BH \gg m_f$,
the particles with high energy $E \gg m_f$ dominate,
then we can approximate
\begin{eqnarray}
 \frac{1}{2} G\left(\frac{m_f}{E}, r\right)
  &\simeq&
 \frac{1}{2} G\left(0, r\right) \nonumber\\
 &=&
 \frac{1}{2}\frac{r^2}{r_\BH^2}
 -
 \frac{1}{4}\frac{r}{r_\BH} \left(3-2\frac{r}{r_\BH}\right)
 \sqrt{1 + 3\frac{r_\BH}{r}}.
\end{eqnarray}
We can perform the $E$-integration in \EQ{mu_eff-2}
because the function $G$ does not depend on $E$ in the limit.
We obtain $H_f(0,r) = \frac{1}{2} G(0,r)$ for bosons and
$H_f(0,r) = \frac{1}{4} G(0,r)$ for fermions.
The effective Higgs mass \EQ{mu_eff-2} becomes
\begin{eqnarray}
 \mu_\eff^2(r)
  &\simeq& -\mu^2
  \;+\;
  \frac{1}{F(r)}
  \frac{A^2}{r^2}
  \;
  \frac{1}{2} G(0, r).
  \label{mu_eff-light}
\end{eqnarray}
We can approximate
\begin{eqnarray}
 \mu_\eff^2(r)
  &\simeq&
  -\mu^2
  \;+\;
  \frac{1}{F(r)}
  \frac{A^2}{r^2}
  \times
  \left\{
   \begin{array}{l@{\qquad}l}
     \frac{27}{16}  & (r \gg r_\BH)\\
     1 & (r \simeq r_\BH)\\
   \end{array}
  \right.
  \label{mu_eff-light-2}
\end{eqnarray}
because $\frac{1}{2} G(0, r)$ is the slowly increasing function
which takes value from $1$ to $\frac{27}{16}=1.6875$.
When we consider the distant region $r \gg r_\BH$,
our high temperature limit \EQ{mu_eff-light-2} agrees
with the non general relativistic limit \EQ{mu_eff-old}
except for the factor $\frac{27}{16}$.
The difference of the factor $\frac{27}{16}$ comes from
the difference of the absorption cross sections,
namely, the gray body factor makes this difference.
When we omit the GR effects for the radiated particle,
the absorption cross section is given by
the horizon area $4\pi r_\BH^2$.
On the other hand,
the absorption cross section for the massless particle
which obeys the Schwarzschild metric
becomes $\sigma_\BH = \frac{27}{4}\pi r_\BH^2$
in \EQ{abs-cross-section}.
When we consider the near-horizon-region,
the effective Higgs mass \EQ{mu_eff-light-2} is governed
by the factor $1/F(r)$ which diverges on the horizon.

{\it Low Temperature Limit} ---
When we consider the black hole
with the low Hawking temperature limit $T_\BH \ll m_f$,
we can approximate
\begin{eqnarray}
 \frac{1}{2} G
  \simeq
 \frac{1}{2} G_{\IIb}
  &=&
  \left\{
   \begin{array}{ll}
     \left(\frac{r}{r_\BH}\right)^2
     \sqrt{1 - \left(\frac{m_f}{E}\right)^2 F(r)} \quad
    & \left( r_\BH \leq r \leq r_\cii\left(\frac{m_f}{E}\right) \right)\\
     0
    & \left( r_\cii\left(\frac{m_f}{E}\right) < r \leq \infty \right)
   \end{array}
  \right..
\end{eqnarray}
Then the effective Higgs mass \EQ{mu_eff-2} becomes
\begin{eqnarray}
 \mu_\eff^2(r)
  &\simeq& -\mu^2
  \;+\;
  \frac{1}{8\pi^2}
  \frac{1}{F(r)}
  T_\BH^2
  \sum_f g_f Y_f^2
  \int_{\frac{m_f\sqrt{F(r)}}{T_\BH}}^\infty ds
   f_{1}(s) \:
   \sqrt{s^2 - \frac{m_f^2 F(r)}{T_\BH^2}}.
  \label{mu_eff-4}
\end{eqnarray}
When we consider the distant region $r \gg r_\cii(\frac{m_f}{T_\BH})$
where the particle with the energy $E \approx T_\BH$ cannot reach,
the only high-energy-side of the tails of the distribution $f_{T_\BH}(E)$
contributes to \EQ{mu_eff-2}.
The difference between boson and fermion is negligible
for \EQ{mu_eff-4} in the distant region
because
the high energy tails of the distribution $f_{T_\BH}(E)$
for bosons and for fermions are almost the same.
Therefore we can adopt the Maxwell distribution
to evaluate \EQ{mu_eff-4} in the distant region
instead of the Fermi-Dirac or the Bose-Einstein distribution.
By using the Maxwell distribution $f_{T_\BH}(E) = \exp(-E/T_\BH)$,
we can perform the integration and obtain
\begin{eqnarray}
 \mu_\eff^2(r)
 &\simeq& -\mu^2
  \;+\;
    \frac{1}{8\pi^2}
     \frac{1}{F(r)} T_\BH^2
    \sum_f g_f Y_f^2 \;
     \frac{m_f\sqrt{F}}{T_\BH} \;
     {\rm K}_1\left(\frac{m_f\sqrt{F}}{T_\BH}\right)
  \label{mu_eff-Maxwell}
\end{eqnarray}
where ${\rm K}_1$ is the modified Bessel function of the second kind.
On the other hand,
when we consider the near horizon region
$r_\BH < r < r_\cii(\frac{m_f}{T_\BH})$,
we can approximate
%
\begin{eqnarray}
  \mu_\eff^2(r)
 &\simeq& -\mu^2
  \;+\;
    \frac{1}{48}
    \frac{1}{F(r)}
    T_\BH^2
    \sum_f \tilde{g}_f(r) Y_f^2,
\end{eqnarray}
where we have defined {\it the local effective degree of the freedom} as
\begin{eqnarray}
 \tilde{g}_f(r) &:=&
  \left\{
   \begin{array}{ll}
    \displaystyle
    \ \ g
    \times \exp\left[
	- 0.60 \sqrt{2}\sqrt{F(r)} \frac{m_f}{T_\BH}
    \right]
    & ({\rm boson})\\[5mm]
    \displaystyle
    \frac{1}{2} g
    \times \exp\left[
	-\left(0.39 \sqrt{2}\sqrt{F(r)} \frac{m_f}{T_\BH}\right)^{3/2}
    \right] \quad
    & ({\rm fermion})
   \end{array}
  \right..
  \label{g_f_r}
\end{eqnarray}
%
%
The only the radiated particle which can reach the position $r$
contributes to the local effective degree of the freedom
$\tilde{g}_f(r)$ on the position $r$.
%

{\it Summarized Form} ---
Finally we can summarize the approximated effective Higgs mass
for all parameter regions as
\begin{eqnarray}
  \mu_\eff^2(r)
 &\simeq& -\mu^2
  \;+\;
    \frac{1}{F(r)}
    \frac{A(r)^2}{r^2},
 \label{mu_eff-final}
\end{eqnarray}
where we have defined {\it the local wall formation parameter} $A(r)$
as the same way of \EQ{wall-formation-constant}
by using the local effective degree of the freedom $\tilde{g}_f(r)$
in \EQ{g_f_r} instead of $\tilde{g}_f$ in \EQ{tilde-g_f}.
As compared with the exact form \EQ{mu_eff-3},
the approximated form \EQ{mu_eff-final} contains an error of the factor one.
By defining {\it the local Hawking temperature} as
\begin{eqnarray}
 \tilde{T}_\BH(r) &:=& \frac{T_\BH}{\sqrt{F(r)}},
\end{eqnarray}
%
%
we can approximate the local effective degree of the freedom \EQ{g_f_r}
as follows
\begin{eqnarray}
 \tilde{g}_f(r) &\simeq&
  \left\{
   \begin{array}{ll}
    \tilde{g}_f \quad & \left( \tilde{T}_\BH(r) \gnear m_f )\right)\\
    0                 & \left( \tilde{T}_\BH(r) \lnear m_f )\right)
   \end{array}
  \right..
  \label{g_f_r-2}
\end{eqnarray}
This property is very similar
to the temperature-dependency of the degree of freedom
in the relativistic thermodynamics.
We present a energy-density-distribution $\rho(r)$ near the horizon
as
$\rho(r)
\;\simeq\;
\frac{\pi^2}{30} g_{*f}(\tilde{T}_\BH(r)) \tilde{T}_\BH^4(r)$
in Appendix B.
This energy-density-distribution reproduces
the relativistic thermodynamical relation
with the temperature $T = \tilde{T}_\BH(r)$.
The region near the horizon is not thermal equilibrium
which is required for thermodynamical treatments,
however,
several behaviors near the horizon are very similar
to that of the thermal equilibrium with
the local Hawking temperature $\tilde{T}_\BH(r)$.

\subsection{Wall Structure Solutions}

\begin{figure}
 \begin{center}
  \includegraphics[scale=1.0]{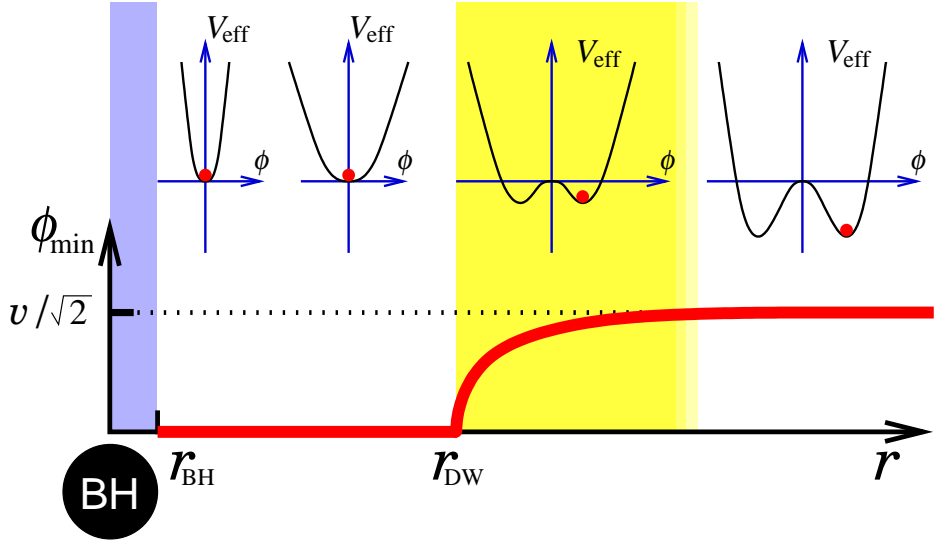}%
 \end{center}
 \caption{%
 Function-forms of the effective Higgs potential $V_\eff(\phi,r)$
 which is depending on the distance from the black hole.
 The dot in the potential-curve means
 the minimum of the potential at each $r$.
 The thick curve is
 the distribution of the Higgs field $\phi_\mini(r)$
 which minimizes the effective potential $V_\eff(\phi,r)$
 at each $r$.
 }%
 \label{WallConcept.eps}
\end{figure}

By substituting the resultant effective Higgs mass \EQ{mu_eff-final},
the effective potential \EQ{Effective-Potential-3} becomes
\begin{eqnarray}
  V_\eff(\phi, r)
  &=&
  \frac{1}{2}
  \left[ -\mu^2 + \frac{1}{F(r)}\frac{A^2(r)}{r^2} \right] \, \phi^2
  \;+\; \frac{1}{2} \frac{\mu^2}{v^2} \, \phi^4.
 \label{Effective-Potential-4}
\end{eqnarray}
The equation for the Higgs vev in \EQ{EOM-phi-3}
with the effective potential \EQ{Effective-Potential-4} becomes
\begin{eqnarray}
 \frac{1}{r^2} \partial_r \left( r^2 F(r) \: \partial_r \phi \right)
  &=&
  \frac{1}{2} \,
  \left( - \mu^2 + \frac{1}{F(r)}\frac{A(r)^2}{r^2} \right) \phi
  \;+\; \frac{\mu^2}{v^2} \, \phi^3.
  \label{EOM-phi-4}
\end{eqnarray}
The factor $1/F(r)$ in
both the effective potential \EQ{Effective-Potential-4}
and the field equation \EQ{EOM-phi-4}
always diverges on the horizon.
The form of the effective potential \EQ{Effective-Potential-4} near the
horizon is different from that in the distant region
(see \fig{WallConcept.eps}).
The minimum of the effective potential
$V_\eff(\phi,r)$ in \EQ{Effective-Potential-4}
depends on the position $r$.
The effective potential $V_\eff(\phi,r)$
is minimized by $\phi=0$ near the horizon
because
the sign of $\mu_\eff^2(r)$ in \EQ{mu_eff-final}
inverts when the position $r$ approach the horizon.
Therefore we expect the formation of the spherical wall-structure
of the Higgs vev around the horizon even if $T_\BH < \mu$.
It is difficult to define the exact radius of the wall-structure
because
the Higgs vev is continuously varying in the wall-structure.
However we can define {\it the characteristic wall radius} $r_\DW$ by
\begin{eqnarray}
 \mu_\eff(r_\DW) = 0,
  \label{def_rDW}
\end{eqnarray}
as a typical scale of the wall-radius.
On the radius the sign of $\mu_\eff^2(r)$ inverts.

\begin{figure}
 \begin{center}
  \includegraphics[scale=1.0]{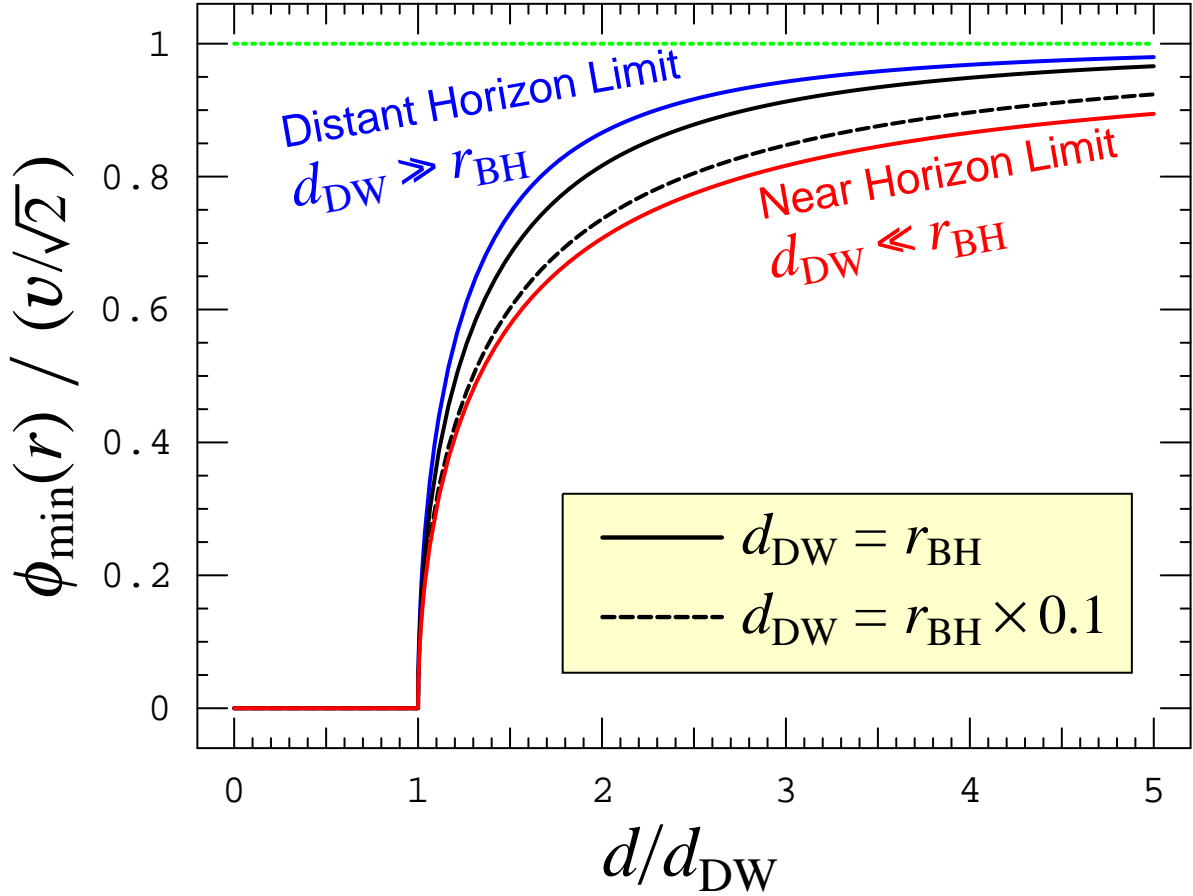}%
 \end{center}
 \caption{%
 Distributions of the Higgs field $\phi_\mini(r)$
 which minimizes the effective potential $V_\eff(\phi,r)$
 at each point.
 The horizontal axis is the distance from the horizon $d$
 normalized by the characteristic-wall-distance $d_\DW$.
 (We note the relations $r = r_\BH + d$ and $r_\DW = r_\BH + d_\DW$.)
 The dotted line $\phi = v/\sqrt{2}$ is the ordinary Higgs vev and
 the vertical axis is normalized by $v/\sqrt{2}$.
 The form of $\phi_\mini(r)$ is depending on the ratio $d_\DW/r_\BH$.
 The uppermost curve describes $\phi_\mini(r)$ in the distant horizon limit
 $d_\DW \rightarrow \infty$ with keeping $r_\BH$ constant.
 The lowest curve describes $\phi_\mini(r)$ in the near horizon limit
 $d_\DW \rightarrow 0$ with keeping $r_\BH$ constant.
 }%
 \label{PhiMin.eps}
\end{figure}

To simplify the analysis of the wall structure
we assume that the wall formation parameter $A^2$ is a constant.
The characteristic wall radius becomes
\begin{eqnarray}
 r_\DW &=& \frac{1}{2}
  \left[
   r_\BH + \sqrt{r_\BH^2 + \frac{A^2}{\mu^2}}
  \right]
  \label{define-r_DW}
\end{eqnarray}
as one of the solutions of the equation \EQ{def_rDW}.
The effective potential \EQ{Effective-Potential-4} is minimized
at each point by the value of the Higgs field:
\begin{eqnarray}
 \phi_\mini(r)
  &=&
  \left\{
   \begin{array}{ll}
    \displaystyle
    \frac{v}{\sqrt{2}}
    \left[ 1 - \frac{F(r_\DW)}{F(r)} \frac{r_\DW^2}{r^2} \right]^{1/2}
      & (r \geq r_\DW) \\[5mm]
    0 & (r < r_\DW)
   \end{array}
  \right.
\end{eqnarray}
The form of $\phi_\mini(r)$ is shown in \fig{PhiMin.eps}
for various values of $d_\DW/r_\BH$.
The normalized form of $\phi_\mini(r)$ is determined by $d_\DW/r_\BH$.

The physical phenomena outside of the horizon are considered
and the inside of the horizon is not relevant to our subject.
To clarify this
we introduce a new positive coordinate $d$ instead of $r$
as a distance from the horizon:
\begin{eqnarray}
 r &=& r_\BH + d.
\end{eqnarray}
We also define a characteristic-wall-distance $d_\DW$
instead of the characteristic-wall-radius $r_\BH$,
which satisfies
\begin{eqnarray}
 r_\DW &=& r_\BH + d_\DW.
\end{eqnarray}
We introduce a profile function of the Higgs vev as
\begin{eqnarray}
 f(s) &:=&
 \frac{1}{v/\sqrt{2}} \phi(r_\BH \;+\; d_\DW \, s),
\end{eqnarray}
which is dimensionless function
and has value $f=1$ in the ordinary vacuum.
The parameter $s$ is a distance from horizon normalized by $d_\DW$.
The equation of the Higgs vev \EQ{EOM-phi-4} becomes
\begin{eqnarray}
   \frac{1}{(s_0 + s)^2}
   \frac{\partial}{\partial s}
   \left[ (s_0 + s) s \; \frac{\partial}{\partial s} f \right]
  &=&
   \frac{A^2}{2}
   \frac{1}{(s_0 + 1)} \,
   \left[
    \left\{ -1 +  \frac{(s_0 + 1)}{(s_0 + s)s} \right\} f
    \;+\; f^3
   \right],
  \label{EOM-phi-5}
\end{eqnarray}
where we have defined a ratio $s_0 := r_\BH/d_\DW$.
This differential equation for the profile function
is depending on both the ratio $s_0$ and the wall-formation constant $A^2$.

To determine the Higgs vev structure around the black hole
by the differential equation \EQ{EOM-phi-5},
we should consider the boundary conditions.
In the ordinary vacuum,
the Higgs vev is $\phi = v/\sqrt{2}$
which is given by the minimum of the bare Higgs potential \EQ{bare-pot}.
The resultant effective potential $V_\eff(\phi,r)$
in \EQ{Effective-Potential-4}
agrees with the bare potential \EQ{bare-pot} for $r\rightarrow\infty$.
Then the boundary condition
\begin{enumerate}
 \item[(a)] $\phi(r) \rightarrow v/\sqrt{2}$ for $r\rightarrow \infty$
\end{enumerate}
is required.
This condition is equivalent to
a condition $f(s) \rightarrow 1$ for $s \rightarrow \infty$.
If there is no Hawking radiation,
we expect that nothing special about the Higgs vev happens on the horizon.
When we turn on the Hawking radiation,
we expect a finite deformation of the Higgs vev around and on the horizon.
Therefore the boundary condition
\begin{enumerate}
 \item[(b)] $\phi(r)$ is finite on the horizon $r=r_\BH$
\end{enumerate}
is also required.
This is equivalent to
the finiteness of the profile function $f(s)$ on $s=0$.

\begin{figure}
 \begin{center}
  \begin{tabular}{c@{\hspace{6mm}}c}
     \includegraphics[scale=0.50]{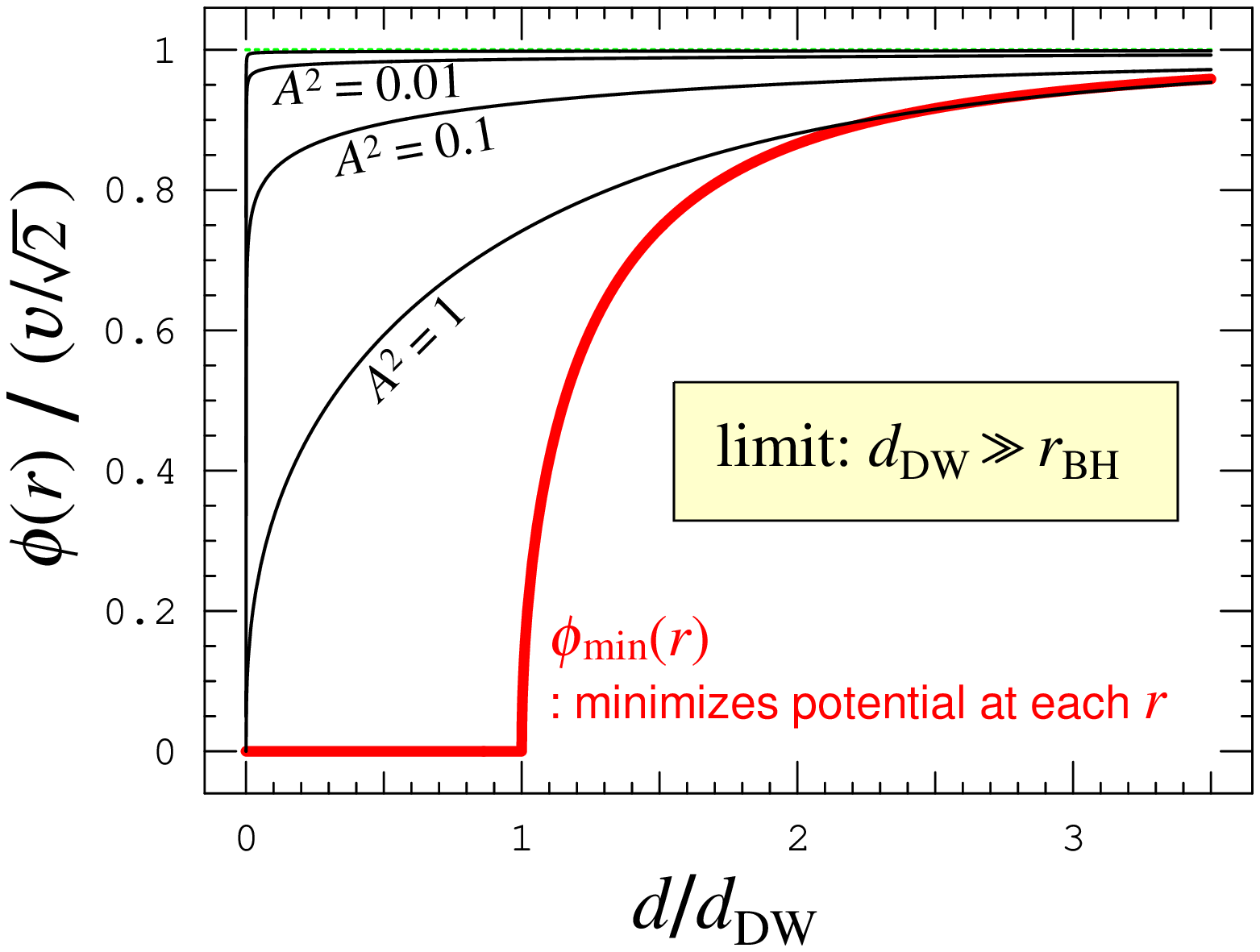}
   & \includegraphics[scale=0.50]{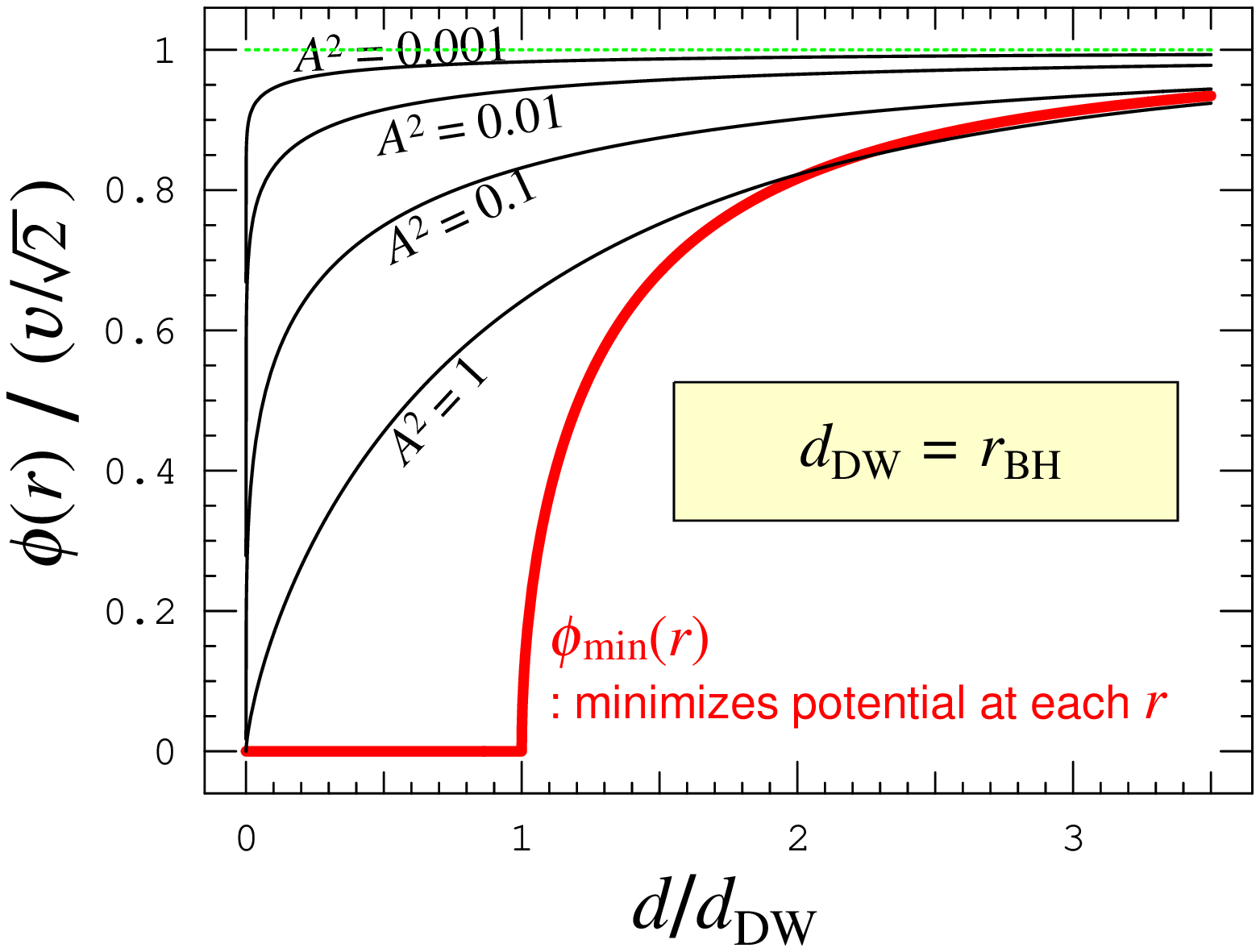}\\
     ({\bf a}) Wall with distant-horizon-limit
   & ({\bf b}) Comparable case of $d_\DW = r_\BH$\\
   \multicolumn{2}{c}{\includegraphics[scale=1.0]{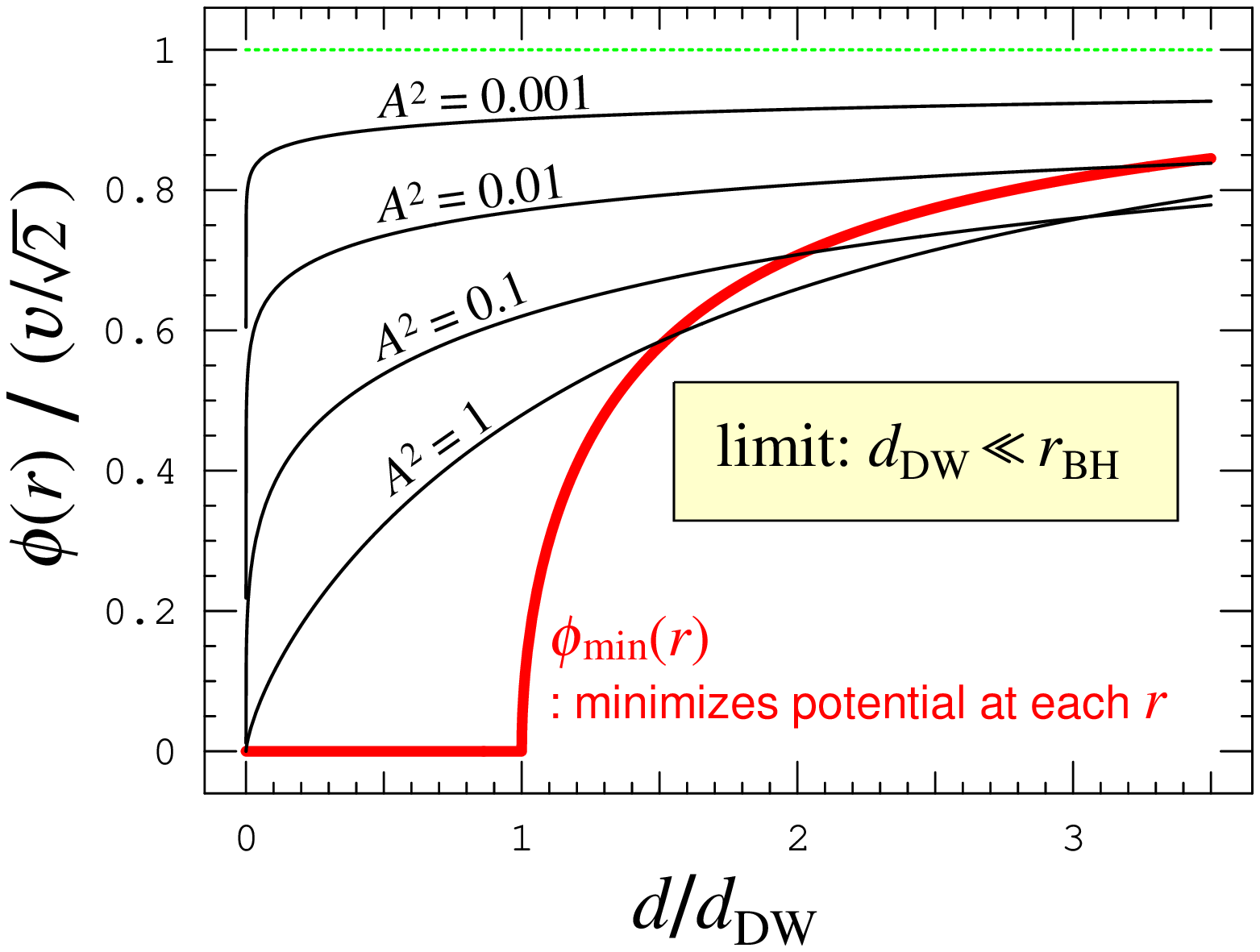}}\\
   \multicolumn{2}{c}{({\bf c}) Wall with near-horizon-limit}
  \end{tabular}
 \end{center}
 \caption{%
 Resultant structures of the Higgs vev around the black hole.
 %
 %
 ({\bf a}) A wall with distant-horizon-limit ($d_\DW \gg r_\BH$),
 ({\bf b}) a comparable case $d_\DW = r_\BH$ and
 ({\bf c}) a wall with near-horizon-limit ($d_\DW \ll r_\BH$)
 are shown.
 For each figure the solutions for
 $A^2 = 0.001, 0.01, 0.1$ and $1$ are shown as thin curves.
 %
 %
 The thick curves $\phi_\mini(r)$ are the Higgs fields
 which minimize the effective potential at each point.
 }%
 \label{WallForm.eps}
\end{figure}

Provided that the length-ratio $s_0 = r_\BH/d_\DW$ and
the wall formation constant $A^2$ are given,
the solution of the differential equation \EQ{EOM-phi-5},
which satisfies the boundary conditions (a) and (b), is uniquely determined.
In the Appendix C
we consider the solutions near the horizon ($d \ll d_\DW$) analytically.
%
%
The near-horizon-solution which satisfy the boundary condition (b)
is given by
\begin{eqnarray}
  f(s) &=&
	C_+ \; s^{+\sqrt{\frac{A^2}{2}}}
	\; \times \nonumber\\
	&& {{}_2F_1} \left[ \textstyle
		\frac{1}{2} + \sqrt{\frac{A^2}{2}}
		- \sqrt{\frac{1}{4} + \frac{A^2}{2}},\;
		\frac{1}{2} + \sqrt{\frac{A^2}{2}}
		+ \sqrt{\frac{1}{4} + \frac{A^2}{2}};\;
		1 + 2\sqrt{\frac{A^2}{2}};\;
		-\frac{s}{s_0}
	\right],
	\label{near-solution}
\end{eqnarray}
where $C_+$ is a positive constant and
${{}_2F_1}$ is the hypergeometric function.
This solution is defined for $s \geq 0$.
The near-horizon-solution \EQ{near-solution}
is a monochromatic increasing function and
has property $f(s) \rightarrow 0$ for $s \rightarrow 0$.
We can numerically solve the equation \EQ{EOM-phi-5}
so that the numerical solution is a numerical extrapolation of
the near-horizon-solution \EQ{near-solution}.
The value of the constant $C_+$ should be so chosen that
the numerical solution satisfies the boundary condition (a).
Namely, the constant $C_+$ is determined
by the matching for the near-horizon-solution and the numerical solutions.
The numerical solutions
for various value of both the parameters $r_\BH/d_\DW$ and $A^2$
are shown in \fig{WallForm.eps}.

The limit $d_\DW \gg r_\BH$ in \fig{WallForm.eps} (a)
is corresponding to the limit $T_\BH \gg \mu$,
because of both the relation $d_\DW \simeq r_\DW \simeq A/\mu$ in this limit
and the relation $T_\BH = \frac{1}{4\pi}\frac{1}{r_\BH}$.
In this limit
the Hawking temperature $T_\BH$ is much greater
than the critical temperature ($\sim\mu$)
of the phase transition of the Higgs field.
The numerical result shown in \fig{WallForm.eps} (a)
indicates the formation of the wall-structure of the Higgs vev.
The general relativistic effects are negligible
because the formed wall-structure is much distant from the horizon.
This result reproduces the wall-structure
concluded in our previous work \cite{BHBG3}
where the general relativistic effects are omitted.

When we consider a comparable case
$d_\DW = r_\BH$ in \fig{WallForm.eps} (b),
the Hawking temperature $T_\BH$ and
the critical temperature ($\sim\mu$) of the gauge-Higgs-Yukawa theory
are also comparable.
We also find the wall-structure around the black hole in the same way.
The Higgs vev vanishes on the horizon: $\phi(r_\BH) = 0$,
then the symmetry which has been spontaneously broken down
by the bare Higgs potential \EQ{bare-pot}
is restored on the horizon.
Therefore the wall separates the symmetric phase region on the horizon
from the background of the broken phase vacuum $\phi = v/\sqrt{2}$.
The restoration of the symmetry on the horizon is
essentially caused by the general relativistic effect
of the Hawking-radiated particles.

The wall-structure for the near-horizon-limit $d_\DW \ll r_\BH$
shown in \fig{WallForm.eps} (c)
is more interesting.
This limit is corresponding to the limit $T_\BH \ll \mu$, namely,
the Hawking temperature is much lower
than the critical temperature ($\sim\mu$) of the phase transition
of the gauge-Higgs-Yukawa theory.
In the previous work \cite{BHBG3}
where the general relativistic effects are omitted,
we concluded that such a cold black hole cannot
form the wall-structure by the Hawking radiation.
However the general relativistic effects of the Hawking radiation
form the wall-structure shown in \fig{WallForm.eps} (c) and
restore the spontaneously-broken-symmetry on the horizon.
The characteristic-wall-distance in this limit becomes
\begin{eqnarray}
 d_\DW &\simeq& \frac{A^2}{4} \frac{1}{r_\BH \mu^2}.
 \label{near-d_DW}
\end{eqnarray}

Finally,
we conclude that the Hawking radiation of a black hole
forms a spherical domain wall around the black hole even
if the Hawking temperature is much smaller
than the critical temperature of the phase transition.

\section{CONCLUSION AND DISCUSSIONS}\label{summary.sec}

In this paper
we have proposed a general relativistic (GR) formulation of the ballistic model
to consider the Hawking radiation from the Schwarzschild black hole
in the vacuum of the gauge-Higgs-Yukawa theory.
We have found that the wall-structure of the Higgs scalar vev is formed
even if the Hawking temperature is smaller than
the energy scale of the gauge-Higgs-Yukawa theory.
In the previous work
where the GR effects were not considered,
the critical Hawking-temperature for the wall-formation
was found out \cite{BHBG3},
however,
we find out the absence of the critical Hawking-temperature.
When the Hawking temperature $T_\BH$ is much smaller than
the energy scale of the field theory $\sim\mu$,
we find the wall-structure closely near the horizon.
The approximated width of the wall-structure near the horizon becomes
the characteristic-wall-distance in \EQ{near-d_DW}.
%
%
On the horizon we also find the restoration of the symmetry
which has been spontaneously broken down
by the bare Higgs potential \EQ{bare-pot}.
These results are essentially produced by the GR effects
on the Hawking-radiated particles and on the Higgs field.

Here we discuss validity of the GR formulation of
the ballistic model.
The ballistic model is valid when
(a) the mean free path of the radiated particles
is longer than the length-scale of the wall-structure $d_\DW$
and
(b) the mean wavelength of the radiation is shorter than the
length-scale $d_\DW$.
The condition (a) is satisfied
when the Hawking temperature $T_\BH$ is smaller than the
critical Hawking temperature of the thermalization $T^*_\BH$
which is much greater
than the critical temperature of the gauge-Higgs-Yukawa theory ($\sim\mu$).
When $T_\BH$ is greater than $T^*_\BH$,
the wall-structure by the thermal phase transition
rather than our mechanism is formed.
This can be confirmed by considering the
interaction-rates around the black hole
\cite{Nagatani:1998gv,Nagatani:2001nz}.
The mean wavelength of the radiated particles
is approximately given by the Schwarzschild radius $r_\BH$
for the region distant from the horizon,
where the GR effects for the radiation are negligible.
When we consider a situation $T_\BH \gnear \mu$,
the wall-structure distant from the horizon ($d_\DW \gnear r_\BH$)
is formed and the condition (b) is satisfied.
On the other hand,
when we consider a situation $T_\BH \lnear \mu$,
the wall-structure near the horizon ($d_\DW \lnear r_\BH$)
is formed and the GR effects become important.
The mean wavelength near the horizon becomes short
by the red-shift effect.
Parikh and Wilczek showed that the Hawking radiation can be regard
as the quantum-tunneling of the particle through the horizon
\cite{Parikh:1999mf}.
They employed a particle description,
which is very similar to our ballistic model,
and the correct spectrum of the Hawking radiation is reproduced.
In their argument the particle description near the horizon becomes valid
because of the red-shift effect.
Then we also expect that our ballistic description is valid
not only for the distant-horizon but also for the near-horizon.

In our GR formulation of the ballistic model,
the Higgs scalar vev around the black hole is determined by
the effective potential $V_\eff(\phi,r; \N)$ in \EQ{Effective-Potential-2}
which is depending on a density-distribution $\N_f(E,r)$
of both the particle-position $r$ and the particle-energy $E$.
The symmetry-restoration on the horizon and 
the wall formation near the horizon
are essentially caused by 
the divergence of the density-distribution $\N_f(E,r)$ on the horizon.
The density-distribution $\N_f(E,r)$ is depending on
the particle-flux $f_f(E,\omega)$ on the horizon,
which is the distribution
of the initial radiation-zenith-angle $\omega$ and
of the particle-energy $E$ (see \fig{ImpactParam.eps}).
The divergence on the horizon of the distribution $\N_f(E,r)$
is caused by the $\cos^{-3}\omega$ factor
of the resultant flux $f_f(E,\omega)$ in \EQ{result-flux}.
The flux $f_f(E,\omega)$ on the horizon is quite different from
that for the normal black-body-radiation \EQ{black-body-flux}
which includes the factor $\cos\omega$.
Therefore domination of the low elevation-angle
(i.e., high initial radiation-zenith-angle $\omega \sim \pi/2$)
on the horizon plays important role
in the wall-formation near the horizon.

Here we summarize the derivation of the flux \EQ{result-flux}
with low elevation-angle domination.
The flux \EQ{result-flux} is derived by
an analytic continuation
of the flux defined for high elevation-angle-region $0 \leq \omega < \omega_c$
into the flux for all elevation-angle $0 \leq \omega < \pi/2$. 
The flux for the high elevation-angle
$0 \leq \omega < \omega_c$ is calculated by the two assumptions;
(i) the motion of the particles radiated from the horizon
is determined by the geodesic on the Schwarzschild space-time
and (ii) the observer at the infinite distance
finds a disk-image with {\it uniform intensity}
of the black-body-radiation for each particle-energy
as the image of the radiated particles.
The assumption (i) is the simplest application
of the general relativity to the Hawking-radiated particles.
The assumption (ii) may be valid
because
the observation of the image with non-uniform strength
means that the Hawking radiation is not thermal radiation
and carries out some information from the black hole.

The domination of the radiation with low elevation-angle
is consistent with the picture of the heat-bath around the horizon
with the local Hawking temperature
$T_\BH(r) \simeq T_\BH / \sqrt{F(r)}$ (see Appendix A).
The energy-density-distribution of the Hawking radiation $\rho(r)$,
which is derived from the resultant particle-density-distribution
$\N_f(E,r)$ in \EQ{N_f-G_f},
is also consistent with $T_\BH(r)$ (see Appendix B).
We have pointed out that $T_\BH(r)$ naturally arises
in the resultant effective Higgs mass in \EQ{mu_eff-final}.
Hotta discussed that
the temperature closely near the horizon becomes very high 
due to the picture of the heat-bath with the local temperature $T_\BH(r)$
and the thermal phase transition
in the string theory arises \cite{Hotta:1997yj}.

The total energy of the Hawking-radiated particle
between the sphere near the horizon with a radius $r_\BH + d$
and the distant sphere with a radius $R$
is given by
\begin{eqnarray}
 E_{\rm total}(d,R) := \int_{r_\BH + d}^{R} 4\pi r^2 dr \rho(r).
\end{eqnarray}
This total energy should be finite, however, this has
an IR divergence for $R\rightarrow\infty$ and
an UV divergence for $d\rightarrow 0$.
Because the black hole has a finite lifetime $\tau_\BH$,
there arises a natural IR cut-off for the radius
$R_{\rm cutoff} \sim \tau_\BH$
Then the IR divergence is easily solved.
On the other hand,
the UV divergence $E_{\rm total} \sim 1/d$ is an open problem.
A cut-off of the Hawking radiation near the horizon may be required.
This cut-off is equivalent to the cut-off for low elevation-angle domination,
namely, the analytic continuation is restricted to
$0 < \omega < \pi/2 - \varepsilon$ with finite $\varepsilon$.
This problem may solved by quantum gravity,
then the cut-off may be given by the Planck length scale
$d_{\rm cutoff} \sim l_\planck$.
If we take the cut-off in the Planck length scale,
our argument in this paper is valid for $d > l_\planck$.
Therefore we expect that our results are valid
as long as we consider low energy ($\ll m_\planck$) phase transition.
Hotta discussed
that the cut-off is given by $T_\BH(r_{\rm cutoff}) \sim m_\planck$
and all of energy and all of the information (entropy)
are carried by the Hawking radiation,
i.e., $E_{\rm total} = m_\BH$  \cite{Hotta:1997yj}.
In this case, the cut-off $r_{\rm cutoff}$ is much smaller than
the Planck length $l_\planck$.

Finally we discuss applications of the wall-structure.
We expect
a net charge-transportation into the black hole by the Hawking radiation
and a phenomenon of the spontaneous charge-up of the black hole
by the effect of the wall-structure \cite{BHBG3}.
We also expect the baryon number production \cite{Sakharov:1967dj} by
the transportation of the hyper charge \cite{Cohen:1991it,Cohen:1993nk}.
This production-mechanism may be realized by
the thin-wall black-hole baryogenesis or
the direct black-hole baryogenesis
proposed in \cite{Nagatani:1998rt}.

The wall-formation near the horizon is caused
by the Hawking-radiated particles with low elevation-angle.
All of such particles return into the horizon and 
are not directly observed by the observer at the infinity distance.
One may consider that
the radiation with low elevation-angle and 
the wall-formation near the horizon
belong to a metaphysical subject.
However the formation of the wall-structure near the horizon
can be confirmed
by the observation of the net charge flux from the black hole.
Therefore the wall-formation near the horizon is not metaphysical.

\begin{flushleft}
 {\Large\bf ACKNOWLEDGMENTS}
\end{flushleft}

 I would like to thank
 Ofer~Aharony, Micha~Berkooz, Alex~Buchel, Hikaru~Kawai, Barak~Kol,
 Joan~Simon and Leonard~Susskind
 for useful discussions.
 I am grateful to K.~Shigetomi
 for helpful advice and also for careful reading of the manuscript.
 The work has been supported by
 the Koshland Postdoctoral Fellowship of the Weizmann Institute of Science.

\section*{APPENDIX A: Radiation Angle Distribution and Metric}

\begin{figure}
 \begin{center}
  \includegraphics[scale=1.0]{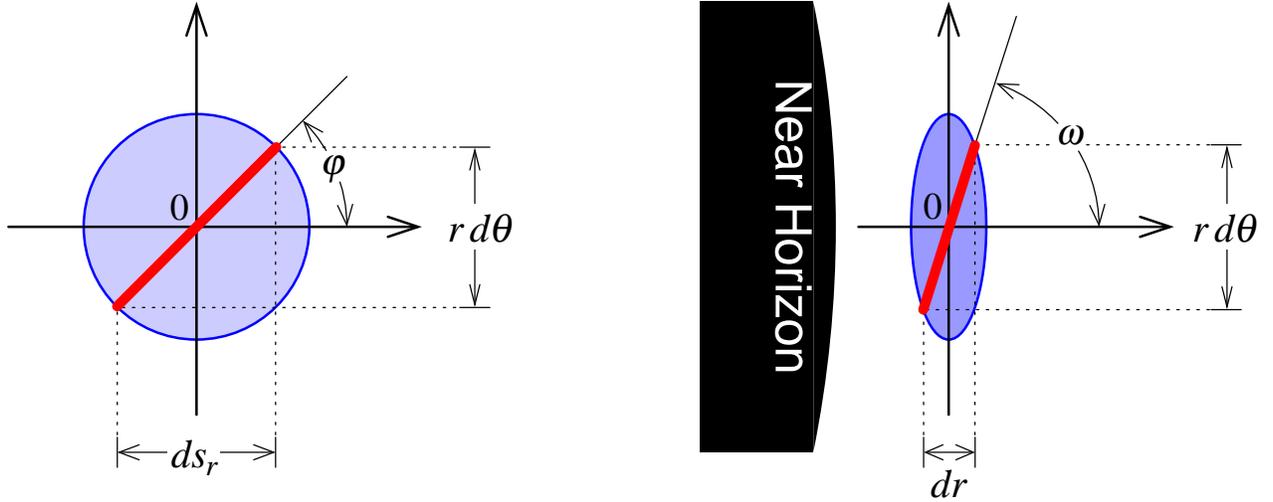}%
 \end{center}
 \caption{%
 (A) A line-element in the proper-length coordinate system
 (thick line segment) and
 (B) the corresponding line-element in the Schwarzschild coordinate system
 near the horizon (thick line segment).
 The relation between the coordinate line-elements is
 $ds_r = F^{-1/2} dr$.
 The definitions of the zenith-angles of the line-element are
 $\tan\varphi = r d\theta/ds_r$ and
 $\tan\omega = r d\theta/dr$.
 Then we have the relation $\tan\varphi = \sqrt{F} \tan\omega$.
 \label{Compress.eps}%
 }%
\end{figure}

In this appendix we show that
the radiation-angle-dependency
of the the Hawking radiation on the horizon
(the factor $1/\cos^{3}\omega$ in \EQ{result-flux})
is naturally derived from the property of the space-time.
Around the black hole described by the Schwarzschild metric \EQ{metric},
a short line-element $dr$ near the horizon
in the Schwarzschild coordinate system
becomes a long line-element
\begin{eqnarray}
 ds_r = \frac{1}{\sqrt{F}} dr.
  \label{ds-dr}
\end{eqnarray}
in the proper coordinate system.
Namely the Schwarzschild coordinate system compress the $r$-direction
near the horizon (see \fig{Compress.eps}).
This compression changes the zenith-angle near the horizon.
The zenith-angle of a line-element
in the the Schwarzschild coordinate system
(the thick segment in \fig{Compress.eps} (B))
is defined as
\begin{eqnarray}
 \tan\omega &=& \frac{r d\theta}{dr}
  \label{A-omega}
\end{eqnarray}
which is used for the radiation-zenith-angle
in the main part of this paper.
On the other hand,
the zenith-angle of the same line-element
in the proper coordinate system
(the thick segment in \fig{Compress.eps} (A))
becomes
\begin{eqnarray}
 \tan\varphi &=& \frac{r d\theta}{ds_r}.
  \label{A-varphi}
\end{eqnarray}
These zenith-angles $\omega$ and $\varphi$ are different description
of the same line-element,
then
a relation between these angles is derived
from the definitions \EQ{A-omega}, \EQ{A-varphi}
and the relation \EQ{ds-dr} as
\begin{eqnarray}
 \tan\varphi &=& \sqrt{F} \tan\omega.
  \label{varphi-omega}
\end{eqnarray}
For convenience we define a function
\begin{eqnarray}
 A(F, \omega) := 1 + \left(\frac{1}{F} - 1\right) \cos^2\omega,
\end{eqnarray}
then we obtain several relations
\begin{eqnarray}
 \frac{\partial\phi}{\partial\omega}
  &=& \frac{1}{\sqrt{F}} \frac{1}{A(F, \omega)},\\
  \cos^2\phi
  &=& \frac{1}{F} \frac{\cos^2\omega}{A(F, \omega)},\\
  \sin^2\phi
  &=& \frac{\sin^2\omega}{A(F, \omega)}.
\end{eqnarray}

We put a angle-distribution of particle-flux
on the $\varphi$-coordinate system as
\begin{eqnarray}
 d\F &=& \tilde{f}(E, \varphi)
  \times dE \times d\varphi \sin\varphi \; d\psi.
  \label{flux-ap}
\end{eqnarray}
The angle-distribution \EQ{flux-ap}
can be rewritten for the $\omega$-coordinate system as
\begin{eqnarray}
 d\F
 &=&
  \tilde{f}(E, \varphi(\omega))
  \frac{d\varphi}{d\omega}
  \frac{\sin\varphi(\omega)}{\sin\omega}
  \times dE \times d\omega \sin\omega \; d\psi \nonumber\\
 &\equiv&
  f(E, \omega)
  \times dE \times d\omega \sin\omega \; d\psi,
\end{eqnarray}
then we obtain the relation
\begin{eqnarray}
  f(E, \omega)
 &=&
  \tilde{f}(E, \varphi(\omega))
  \frac{1}{F^{1/2} A(F,\omega)^{3/2}},
\end{eqnarray}
which gives us the coordinate transformation law
$(\omega \leftrightarrow \varphi)$
for the angle-distribution.

When we consider the near-horizon-limit $r\rightarrow r_\BH$,
we have $F(r) \rightarrow 0$ and
the coordinate transformation for the zenith-angles
\EQ{varphi-omega} becomes
\begin{eqnarray}
 \varphi(\omega)
  &\rightarrow&
  \left\{
 \begin{array}{l@{\qquad}l}
  0     & (0 \leq \omega \lnear \pi/2) \\
  \pi/2 & (\omega = \pi/2)
 \end{array}
 \right..
\end{eqnarray}
We also obtain the following forms in the limit:
\begin{eqnarray}
 A(F, \omega) &\rightarrow& \frac{1}{F} \cos^2\omega,\\
  f(E, \omega) &\rightarrow& 
  \tilde{f}(E, \varphi(\omega))
  \frac{F}{\cos^3\omega}.
\end{eqnarray}

If we assume that
the zenith-angle-distribution of the flux $\tilde{f}(E, \varphi)$
in the proper coordinate system is smooth around $\varphi = 0$,
we have 
\begin{eqnarray}
  f(E, \omega) &=& 
  \tilde{f}(E, 0)
  \frac{F}{\cos^3\omega}.
  \label{result-flux-appendix}
\end{eqnarray}
in the near-horizon-limit.
Here we find another derivation of the factor $1/\cos^{3}\omega$
in the angle-distribution of the particle-flux \EQ{result-flux-appendix}.
In Section \ref{angle.sec}
we have derived the distribution of the particle-flux
of the Hawking radiation in \EQ{result-flux}
including the factor $1/\cos^{3}\omega$
by analytic continuation and uniformity of the Hawking radiation.
On the other hand, the factor $1/\cos^{3}\omega$
is derived from the smoothness of the distribution
and geometry of the black hole in this appendix.

Inversely if we assume the resultant flux \EQ{result-flux},
we obtain the flux on the proper coordinate system as
\begin{eqnarray}
 \tilde{f}_f(E, 0)
  &=&
  \frac{1}{16\pi} \frac{1}{F} \times
  g_{f} f_{T_\BH}(E) \: 4\pi E^2 \nonumber\\
  &=&
  \frac{1}{4\pi} \times
  g_{f} f_{\tilde{T}_\BH}(\tilde{E}) \: 4\pi \tilde{E}^2,
  \label{proper-flux-ap}
\end{eqnarray}
where we have defined
the proper energy of the particle $\tilde{E}(r) := E / \sqrt{2 F(r)}$ and
the effective local Hawking temperature
$\tilde{T}_\BH(r) := T_\BH / \sqrt{2 F(r)}$.
The proper flux for the zenith-direction $\varphi \simeq 0$
resulted in \EQ{proper-flux-ap}
is consistent with
the heat-bath picture with the local Hawking temperature $\tilde{T}_\BH(r)$.

\section*{APPENDIX B: Energy Density Distribution around the Black Hole}

\begin{figure}
 \begin{center}
  \begin{tabular}{c@{\hspace{6mm}}c}
     \includegraphics[scale=0.47]{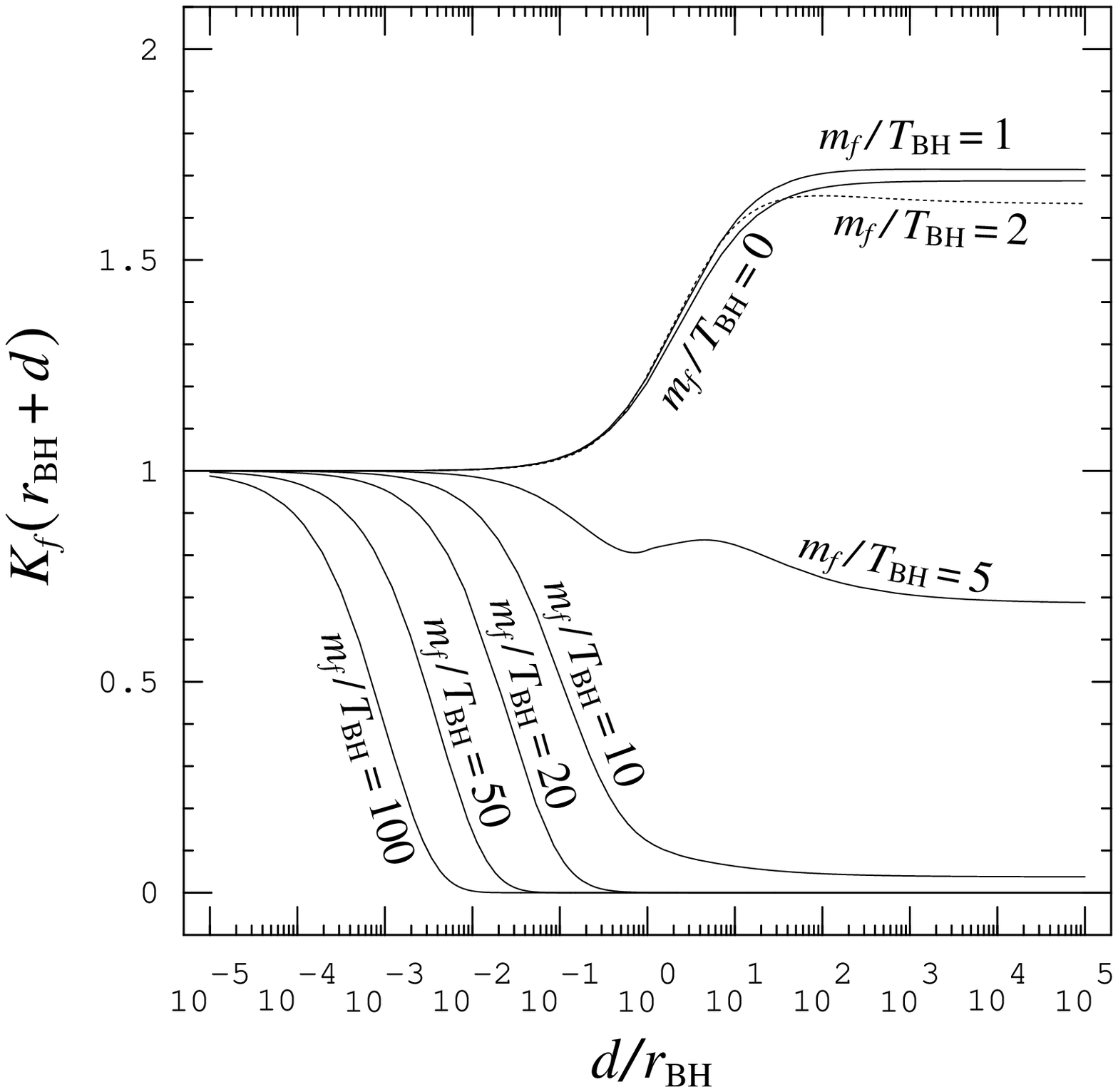}
   & \includegraphics[scale=0.47]{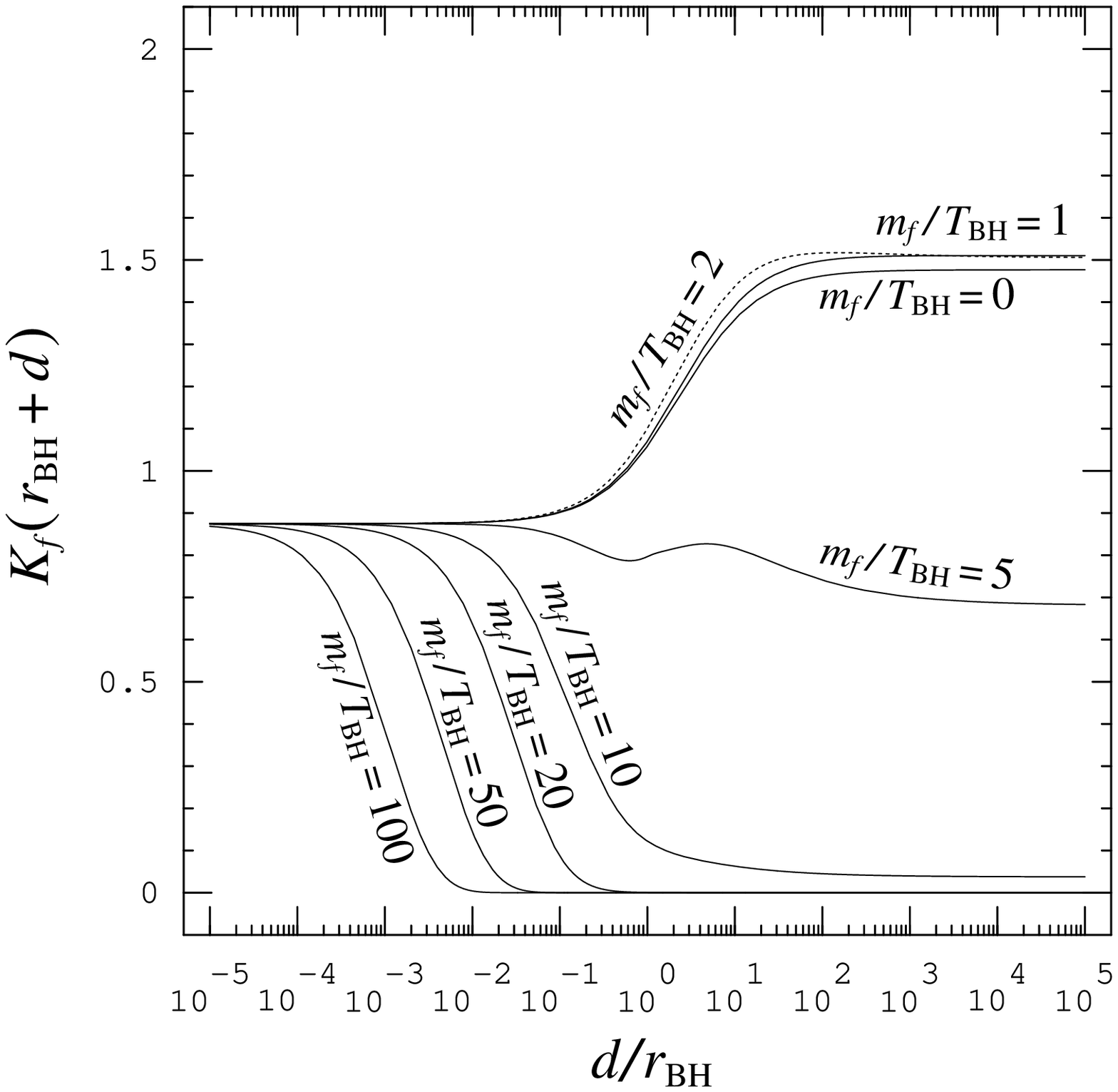}\\
     ({\bf a}) Boson
   & ({\bf b}) Fermion
  \end{tabular}
 \end{center}
 \caption{%
 The shapes of the form-function $K_f(\frac{m_f}{T_\BH},r)$
 defined in \EQ{K_f}
 for various $m_f/T_\BH$ and
 for ({\bf a}) bosons and ({\bf b}) fermions.
 The horizontal axis $d$ is the distance from the horizon
 of the black hole with the Schwarzschild radius $r_\BH$.
 \label{Kf.eps}%
 }%
\end{figure}

In this appendix
we calculate the energy density of the Hawking-radiated particles
around the black hole
to clarify the physical meaning of our resultant
particle number-distribution $N_f(E,r)$ in \EQ{N_f-G_f}.
The energy density of the Hawking-radiated particle
at the position $r$ with the particle-species $f$ is given by 
\begin{eqnarray}
 \rho_f(r) &=& \int_0^\infty dE \: E \: \N_f(E,r).
 \label{rho-app-1}
\end{eqnarray}
By defining the form function
\begin{eqnarray}
 K_f\left(\frac{m_f}{T_\BH},r\right)
  &:=&
  \frac{15}{\pi^4} \int_0^\infty ds \; s^3 \: f_1(s)
  \frac{1}{2} G \left(\frac{m_f}{T_\BH}\frac{1}{s},r\right),
  \label{K_f}
\end{eqnarray}
we obtain
\begin{eqnarray}
 \rho_f(r)
  &=& \frac{\pi^2}{120} g_f T_\BH^4 
   \; \times \; 
   \frac{1}{F^2(r)}
   \left(\frac{r_\BH}{r}\right)^2
   K_f\left(\frac{m_f}{T_\BH},r\right).
   \label{rho-app-2}
\end{eqnarray}
We display the shapes of the form function $K_f(r)$ in \fig{Kf.eps}.
For any $m_f/T_\BH$, we have the same near-horizon limits
$K_f(r\rightarrow r_\BH) \rightarrow 1$ for bosons and
$K_f(r\rightarrow r_\BH) \rightarrow 7/8$  for fermions.
For the massless particle we have
$K_f(r\rightarrow \infty) \rightarrow \frac{27}{16}\simeq1.69$
for bosons and
$K_f(r\rightarrow \infty) \rightarrow
\frac{27}{32}\times\frac{7}{8}\simeq1.48$
for fermions.

The energy-density \EQ{rho-app-2} has an interesting approximated form:
\begin{eqnarray}
 \rho_f(r)
  &\simeq&
  \frac{\pi^2}{30} g_{*f}(\tilde{T}_\BH(r)) \; \tilde{T}_\BH^4(r)
  \;\times\; \left(\frac{r_\BH}{r}\right)^2,
  \label{rho-app-3}
\end{eqnarray}
where we have defined the local Hawking temperature
\begin{eqnarray}
 \tilde{T}_\BH(r) &:=& \frac{1}{\sqrt{2}} \frac{T_\BH}{\sqrt{F(r)}}
\end{eqnarray}
and also defined the degree of freedom of the particle $f$
\begin{eqnarray}
 g_{*f}(T) &:=&
  \left\{
   \begin{array}{ll}
    g_f(T) & (f : {\rm boson})\\
    \frac{7}{8} g_f(T) \quad & (f : {\rm fermion})
   \end{array}
  \right.
\end{eqnarray}
for the temperature $T$ with the fermion-correction.
The relation \EQ{rho-app-3} is just
the thermodynamical relations
between the temperature $\tilde{T}_\BH(r)$
and the energy density $\rho_f(r)$
except for the spherical geometrical factor $(r_\BH/r)^2$.
Then our evaluation of the differential particle-number-density \EQ{N_f-G_f}
is consistent with the picture of a heat-bath around a black hole
with the local temperature $\tilde{T}(r)$.

\section*{APPENDIX C: Solution near Horizon}

We consider the properties of the solution
closely near the horizon: $r=r_\BH + d$ with $d \ll r_\BH$.
In the region near the horizon
the positive contribution to the Higgs mass term
dominates in the effective potential \EQ{Effective-Potential-4} 
because
the Schwarzschild factor $\frac{1}{F(r)}$ becomes very large
near the horizon\footnote{
The reason can be regard that
the local Hawking temperature $T_\BH/\sqrt{F(r)}$ becomes very large.}.
The field equation \EQ{EOM-phi-5} for the profile function
$f(s) = \frac{1}{v/\sqrt{2}} \phi(r_\BH \;+\; d_\DW \, s)$
is approximated to the linear differential equation:
\begin{eqnarray}
   \frac{1}{(s_0 + s)}
   \frac{\partial}{\partial s}
   \left[ (s_0 + s) s \; \frac{\partial}{\partial s} f \right]
  &=&
   \frac{A^2}{2}
   \frac{1}{s} \; f,
  \label{EOM-phi-5-A}
\end{eqnarray}
where we had defined the ratio $s_0 = r_\BH/d_\DW$.
The general solution of this second order differential equation is given by
\begin{eqnarray}
 f(s) &=&
	C_+ \; s^{+\sqrt{\frac{A^2}{2}}} \; \times \nonumber\\
	&& {{}_2F_1} \left[ \textstyle
		\frac{1}{2} + \sqrt{\frac{A^2}{2}}
		- \sqrt{\frac{1}{4} + \frac{A^2}{2}},\;
		\frac{1}{2} + \sqrt{\frac{A^2}{2}}
		+ \sqrt{\frac{1}{4} + \frac{A^2}{2}},\;
		1 + 2\sqrt{\frac{A^2}{2}},\;
		-\frac{s}{s_0}
	\right]
	\nonumber\\
	&+&
	C_- \; s^{-\sqrt{\frac{A^2}{2}}} \; \times \nonumber\\
	&& {{}_2F_1} \left[ \textstyle
		\frac{1}{2} - \sqrt{\frac{A^2}{2}}
		- \sqrt{\frac{1}{4} + \frac{A^2}{2}},\;
		\frac{1}{2} - \sqrt{\frac{A^2}{2}}
		+ \sqrt{\frac{1}{4} + \frac{A^2}{2}},\;
		1 - 2\sqrt{\frac{A^2}{2}},\;
		-\frac{s}{s_0}
	\right],
	\label{approx-sol-a1}
\end{eqnarray}
where ${{}_2F_1}$ is the hypergeometric function.
The coefficients $C_+$ and $C_-$ are determined by the boundary conditions.
The first term is a monochromatic increasing function and has zero value
on the horizon ($s=0$).
The second term has singularity on the horizon.
To obtain the Higgs vev by the differential equation \EQ{EOM-phi-5-A}
we have required the boundary conditions;
(a) $f(s) \rightarrow 1$ for $s \rightarrow \infty$ and
(b) $f(s)$ is finite for $s \rightarrow 0$.
Due to the boundary condition (b)
the second coefficient $C_-$ should be zero.
The first coefficient $C_+$ is determined by
the matching of the solution \EQ{approx-sol-a1} with $C_- = 0$
and the numerical solution which satisfies the boundary condition (a).
As a result of matching to the numerical solutions
shown in \fig{WallForm.eps},
we obtain $O(1)$ values of $C_+$.
For example we obtain
$C_+ = 0.90$ for $A^2 = 0.001$ with the wall extremely near the horizon,
$C_+ = 1.00$ for $A^2 = 0.001$ with $d_\DW = r_\BH$
and 
$C_+ = 1.14$ for $A^2 = 0.001$ with larger wall than the scale of the horizon.

The analysis of the Higgs vev near the horizon
results that
the symmetry broken-down spontaneously by the bare Higgs potential
is restored on the horizon $f(0) = 0$
and that
the spherical wall-structure of the Higgs vev separates
the symmetric phase region on the horizon
from the broken phase vacuum in the region distant from the horizon.
Therefore the resultant wall-structure 
is just the ``domain wall'',
according to the ordinary terminology.

To clarify the behavior of the solution \EQ{approx-sol-a1},
we will consider the extreme cases.
When we consider the wall with the near horizon limit, i.e.,
$s_0 = r_\BH/d_\DW \rightarrow \infty$,
the solution becomes a simple form:
\begin{eqnarray}
 f(s) &=&
  C_+ \; s^{+\sqrt{\frac{A^2}{2}}}
  \;+\;
  C_- \; s^{-\sqrt{\frac{A^2}{2}}}.
\end{eqnarray}
When we consider the wall which is extremely distant from the horizon,
i.e., $s_0 = r_\BH/d_\DW \rightarrow 0$,
the solution becomes
\begin{eqnarray}
 f(s) &=&
    C_+ \; s^{-\frac{1}{2} + \sqrt{\frac{1}{4} + \frac{A^2}{2}}}
    \;+\;
    C_- \; s^{-\frac{1}{2} - \sqrt{\frac{1}{4} + \frac{A^2}{2}}}.
\end{eqnarray}
In both cases, $C_- = 0$ is required by the boundary condition (b).
When the wall-formation-constant is much small, i.e., $A^2 \ll 1$,
the solution near the horizon which satisfies the boundary condition (b)
can be summarized as
\begin{eqnarray}
  f(s) &\simeq& C_+ \; s^{+\sqrt{\frac{A^2}{2}}}.
\end{eqnarray}
When we consider the electroweak wall-structure,
we have the wall-formation-constant $A^2 \simeq 0.001$ \cite{BHBG3}
then we can employ this form.

\newcommand{\PRL}[3]	{{Phys.\ Rev.\ Lett.}   {\bf #1}, #2 (#3)}
\newcommand{\PR}[3]	{{Phys.\ Rev.}          {\bf #1}, #2 (#3)}
\newcommand{\PRA}[3]	{{Phys.\ Rev.\ A}       {\bf #1}, #2 (#3)}
\newcommand{\PRD}[3]	{{Phys.\ Rev.\ D}       {\bf #1}, #2 (#3)}
\newcommand{\PL}[3]	{{Phys.\ Lett.}         {\bf #1}, #2 (#3)}
\newcommand{\PLA}[3]	{{Phys.\ Lett.\ A}      {\bf #1}, #2 (#3)}
\newcommand{\PLB}[3]	{{Phys.\ Lett.\ B}      {\bf #1}, #2 (#3)}
\newcommand{\NuP}[3]	{{Nucl.\ Phys.}         {\bf #1}, #2 (#3)}
\newcommand{\PTP}[3]	{{Prog.\ Theor.\ Phys.} {\bf #1}, #2 (#3)}
\newcommand{\Nature}[3]	{{Nature}               {\bf #1}, #2 (#3)}
\newcommand{\PKNAW}[3]	{{Proc.\ K.\ Ned.\ Akad.\ Wet.} {\bf #1}, #2 (#3)}
\newcommand{\Physica}[3]{{Physica\ (Utrecht)}   {\bf #1}, #2 (#3)}
\newcommand{\JMP}[3]	{{J.\ Math.\ Phys.}     {\bf #1}, #2 (#3)}
\newcommand{\PRSLA}[3]	{{Proc.\ R.\ Soc.\ London,\ Ser A} {\bf #1}, #2 (#3)}
\newcommand{\AP}[3]	{{Ann.\ Phys.\ (N.Y.)}  {\bf #1}, #2 (#3)}
\newcommand{\JPA}[3]	{{J.\ Phys.\ A}         {\bf #1}, #2 (#3)}
\newcommand{\ZhETF}[3]	{{Zh.\ \'{E}ksp.\ Teor.\ Fiz.\ Pis'ma.\ Red.} {\bf #1}, #2 (#3)}
\newcommand{\JETP}[3]	{{JETP\ Lett.}          {\bf #1}, #2 (#3)}
\newcommand{\CMP}[3]	{{Commun.\ Math.\ Phys.}{\bf #1}, #2 (#3)}
\newcommand{\MPLA}[3]   {{Mod.\ Phys.\ Lett.}   {\bf A#1}, #2 (#3)} 


\end{document}